\documentclass[twocolumn]{aastex63}

\usepackage{apjfonts}

\def\d3{$\delta_{3}$ }
\def\1d3{$(1 + \delta_{3})$ }
\def\l1d3{$\log_{10}(1 + \delta_{3})$ }

\def\s3{$\Sigma_{3}$}
\def\ha{H$\alpha$}
\def\hb{H$\beta$}

\def\24m{24 $\mu$m}

\def\sm{$\rm~M_{*}$}

\def\kms{${\rm km~s^{-1}}$ }

\def\Msolar{$\rm M_{\odot}$}

\def\rmxaa{RMxAA}

\def\sigsm{$\Sigma_{*}$}
\def\sigsfr{$\Sigma_{\rm SFR}$}

\def\h2{$\rm H_{2}$}
\def\Mh2{$\rm M_{H_{2}}$}

\def\sigh2{$\Sigma_{\rm H_{2}}$}
\def\fgas{$f_{\rm gas}$}
\def\fh2{$f_{\rm H_{2}}$}

\def\co{$^{12}$CO(1-0)}

\shorttitle{Molecular Gas in Green Valley Galaxies}
\shortauthors{Lin et al.}

\begin{document}

\title{ALM\lowercase{a}QUEST - VII: Star Formation Scaling Relations of Green Valley Galaxies}

\author{Lihwai Lin}
\altaffiliation{Email: lihwailin@asiaa.sinica.edu.tw}
\affiliation{Institute of Astronomy \& Astrophysics, Academia Sinica, Taipei 10617, Taiwan}

\author{Sara L. Ellison}
\affiliation{Department of Physics \& Astronomy, University of Victoria, Finnerty Road, Victoria, British Columbia, V8P 1A1, Canada}

\author{Hsi-An Pan}
\affiliation{Max-Planck-Institut f\"ur Astronomie, K\"onigstuhl 17, D-69117 Heidelberg, Germany}
\affiliation{Department of Physics, Tamkang University, No.151, Yingzhuan Rd., Tamsui Dist., New Taipei City 251301, Taiwan}

\author{Mallory D. Thorp}
\affiliation{Department of Physics \& Astronomy, University of Victoria, Finnerty Road, Victoria, British Columbia, V8P 1A1, Canada}

\author{Po-Chieh Yu}
\affiliation{College of General Studies, Yuan-Ze University, Taoyuan 32003, Taiwan}

\author{Francesco Belfiore}
\affiliation{INAF - Osservatorio Astrofisico di Arcetri, Largo E. Fermi 5, I-50157 Firenze, Italy}

\author{Bau-Ching Hsieh}
\affiliation{Institute of Astronomy \& Astrophysics, Academia Sinica, Taipei 10617, Taiwan}

\author{Roberto Maiolino}
\affiliation{Cavendish Laboratory, University of Cambridge, 19 J. J. Thomson Avenue, Cambridge CB3 0HE, United Kingdom}
\affiliation{University of Cambridge, Kavli Institute for Cosmology, Cambridge, CB3 0HE, UK}

\author{S. Ramya}
\affiliation{Indian Institute of Astrophysics, II Block, Koramangala, Bengaluru 560 034, India}

\author{Sebasti\'{a}n F. S\'{a}nchez }
\affiliation{Instituto de Astronom\'ia, Universidad Nacional Aut\'onoma de  M\'exico, Circuito Exterior, Ciudad Universitaria, Ciudad de M\'exico 04510, Mexico}

\author{Yung-Chau Su}
\affiliation{Institute of Astronomy \& Astrophysics, Academia Sinica, Taipei 10617, Taiwan}
\affiliation{Department of Physics, National Taiwan University, 10617, Taipei, Taiwan}

\begin{abstract}

We utilize the ALMA-MaNGA QUEnch and STar formation (ALMaQUEST) survey to investigate the kpc-scale scaling relations, presented as the resolved star forming main sequence (rSFMS: \sigsfr~vs. \sigsm), the resolved Schmidt-Kennicutt relation (rSK: \sigsfr~vs. \sigh2), and the resolved molecular gas main sequence (rMGMS: \sigh2~vs. \sigsm), for 11478 star-forming and 1414 retired spaxels (oversampled by a factor of $\sim20$) located in 22 green valley (GV) and 12 main sequence (MS) galaxies.  For a given galaxy type (MS or GV), the retired spaxels are found to be offset from the sequences formed by the star-forming spaxels on the rSFMS, rSK, and rMGMS planes, toward lower absolute values of sSFR, SFE, and \fh2~ by $\sim$ 1.1, 0.6, and 0.5  dex. The scaling relations for GV galaxies are found to be distinct from that of the MS galaxies, even if the analyses are restricted to the star-forming spaxels only. It is found that for star-forming spaxels, sSFR, SFE, and \fh2~in GV galaxies are reduced by $\sim$0.36, 0.14, and 0.21 dex, respectively, compared to those in MS galaxies. Therefore, the suppressed sSFR/SFE/\fgas~in GV galaxies is associated with not only an increased proportion of retired regions in GV galaxies but also a depletion of these quantities in star-forming regions. Finally, the reduction of SFE and \fh2~in GV galaxies relative to MS galaxies is seen in both bulge and disk regions (albeit with larger uncertainties), suggesting that statistically, quenching in the GV population may persist from the inner to the outer regions.

\end{abstract}

\keywords{galaxies:evolution $-$ galaxies: low-redshift $-$ galaxies: star formation $-$ galaxies: ISM}

\section{INTRODUCTION}

One of the central keys in probing galaxy evolution lies in the understanding of how the star formation in galaxies is regulated, triggered, and suppressed. All of these processes are closely linked to the amount of the molecular gas that is available to fuel star formation, conventionally parameterized by the gas fraction, and the physical conditions of the gas that determines whether the gas is able to collapse to form stars, represented by the star formation efficiency \citep[e.g.,][]{won02,big08,sar14,sai17,tac18,san20}. 

With the advent of Integral Field Unit (IFU) surveys and high spatial resolution radio/mm observations, the connection between star formation, stellar properties, and molecular gas contents  have been explored recently at (sub)kpc scales with statistical samples \citep[e.g.,][]{lin19b,ell20a,ell20b,bro20,mor20,san20,san21,bar21,ell21a,ell21b,pes21}. Using 14 galaxies from the ALMA-MaNGA QUEnching and STar formation (ALMaQUEST) survey \citep{lin20}, \citet{lin19b} established the scaling relations among the star formation rate surface density (\sigsfr), stellar mass surface density (\sigsm), and \h2~mass surface density (\sigh2) for star-forming spaxels of main sequence galaxies on kpc scales. The three variables are found to form a 3D linear (in log) relation with associated scatter. As a result, each pair of these quantities displays a tight sequence, dubbed the resolved star-forming main sequence (rSFMS: \sigsfr~ vs. \sigsm), the resolved Schmidt-Kennicutt relation (rSK: \sigsfr~ vs. \sigh2), and the resolved molecular gas main sequence (rMGMS: \sigh2~ vs. \sigsm). Similar results have also been obtained with a larger set of ALMaQUEST galaxies \citep{ell20b}, EDGE-CALIFA \citep{san20,san21}, PHANGS \citep{pes21}, and other nearby galaxies \citep{mor20}. These works have led to several interesting findings. For instance, the kpc-scale relations are found to resemble the global relations, implying that the global scaling relations could be originated from local processes at sub-galactic scales \citep{san13,wuy13,can16,hsi17,abd17,pan18a,lin19b,can19,vul19,eni20,san20}. Studies of the scatters, correlation coefficients, galaxy-to-galaxy variation, as well as the co-variance of the these scaling relations suggest that rSFMS (and hence the global SFMS) could simply be a consequence of the combination of rSK and rMGMS \citep{lin19b,ell21a,mor20,pes21,bak21}. Moreover, \citet{ell21a} find significant correlations between the scaling relations and properties such as Sersic index and total stellar mass. On the other hand, \citet{san21} explore the correlations between the residuals of the three scaling relations in the EDGE-CALIFA sample and find that the scatter of the 3D scaling relation is dominated by the measurement errors and hence does not have a physical origin. Other explanations also include the hydrostatic mid-plane pressure being the primary driver in regulating local star formation given its strong correlation with \sigsfr\citep{bar21}.

Despite the detailed characterization of star-forming spaxels in the aforementioned studies, most previous work has focused on galaxies that are still actively forming stars (i.e., past studies have restricted themselves to \textit{star-forming} spaxels in \textit{star-forming} galaxies). Whether the resolved scaling relations still hold or not in galaxies departing from the main sequence, such as the green valley (GV) or quiescent populations, is less explored. Furthermore, it is also less clear how the non-star-forming spaxels behave, particularly for the quenched or `retired' regions where the star formation has already ceased or is largely suppressed. The `retired' regions are conventionally identified as those having \ha~ EW $<$ 3 \AA~ \citep{sta08,cid10,san14,ell21b}. Under this definition, they are often referred to areas whose emission lines are primarily photoionized by aging stellar populations rather than newly formed stars, even though they could actually have just as much star formation as some spaxels in regions with fewer old stars present to add ionization. \citet{ell21b} undertook a first initiative to investigate the molecular gas fraction in the retired spaxels using 8 of the ALMaQUEST galaxies and found that in a given galaxy, the retired spaxels form a rMGMS sequence that is several times lower compared to that of the star-forming spaxels, indicating lower gas fractions in retired regions.  The distinction between gas fractions in star-forming and retired spaxels is seen even within a given galaxy.

The ratios of the two variables in the rSFMS, rSK, and rMGMS represent the three normalized quantities, the specific star formation rate (sSFR $\equiv$ \sigsfr/\sigsm), star formation efficiency (SFE $\equiv$ \sigsfr/\sigh2), and molecular gas fraction (\fh2 $\equiv$ \sigh2/\sigsm), respectively. As the rate of forming stars depends on both the amount of molecular gas and the efficiency of converting molecular gas into stars such that sSFR = SFE $\times$ \fh2, characterizing the scaling relations in retired areas or quenched galaxies can shed light on the relative contributions between SFE and \fh2~to sSFR in quenching star formation. With the \co~observations obtained from the Atacama Large Millimeter Array (ALMA) of three Mapping Nearby Galaxies at Apache Point Observatory \citep[MaNGA;][]{bun15} selected green valley galaxies, \citet{lin17} reported a stronger depletion in \fh2~ in the bulges compared to the disk regions of green valley galaxies \citep[also see][]{san18,bro20}, consistent with an inside-out quenching scenario \citep{bel18,ell18,lin19a}. However, a larger sample is required to confirm the trend.

In this work, we extend the study of the star formation scaling relation carried out by \citet{lin19b}, which was based on well-selected star-forming spaxels of the main sequence population, to retired regions, as well as comparing scaling relations of green valley galaxies with main sequence galaxies in the ALMaQUEST sample. We aim at characterizing and comparing the scaling relations between different types of spaxels (retired vs. star-forming) and galaxies (main sequence vs. green valley) in order to gain a full picture of the connection between star formation and gas properties at kpc scales. In addition, we also quantify the suppression of SFE and \fh2~in bulges and disks of green valley galaxies separately, to investigate whether there is a preferential location within galaxies for the quenching to start.

Throughout this paper we adopt the following cosmology: \textit{H}$_0$ = 70~\kms Mpc$^{-1}$, $\Omega_{\rm m} = 0.3$ and $\Omega_{\Lambda } = 0.7$. We use a Salpeter initial mass function (IMF).

\section{SAMPLE and OBSERVATIONS \label{sec:data}}

\subsection{The ALMaQUEST survey}
The ALMaQUEST survey \citep{lin20} maps the \co~distributions with ALMA for 46 galaxies selected from the MaNGA
survey. The sample consists of galaxies with a wide range of specific star formation rates (sSFR), including starburst (SB), main sequence (MS), and green valley (GV) galaxies. The ALMA observations were carried out with a spatial resolution that is matched to the point spread function ($\sim 2.5$ arcsec) of MaNGA, enabling a joint study of stellar and gas properties at the same physical scales (corresponding to a physical scale of 0.9--6 kpc). The details of the sample characteristics and data reduction are given in the ALMaQUEST survey paper \citep{lin20}. In this work, the H$_{2}$ mass surface density, \sigh2, is computed from the CO luminosity by adopting a constant conversion factor ($\alpha_{\mathrm{CO}}$) of 4.35 \Msolar (K km s$^{-1}$ pc$^{2}$)$^{-1}$ \citep[e.g.,][]{bol13}. 

We can first compare the ALMaQUEST sample with other spatially resolved IFS and molecular gas observations in nearby galaxies. For example, the PHANGS-MUSE survey (PI: E. Schinnerer; Emsellem et al. in prep.) offers resolved observations with $>$ ten times better spatial resolution, down to 50-100 pc for 19 star-forming disk galaxies selected from the PHANGS-ALMA survey \citep{ler21}. However, the PHANGS-MUSE survey only targets galaxies on the star formation main sequence and therefore does not sample galaxies going through the quenching phase (e.g., green valley) as those included in the ALMaQUEST survey. On the other hand, the EDGE-CALIFA survey \citep{bol17} consists of a larger size of sample (126 galaxies) with resolved CO observations selected from the CALIFA \citep{san12} survey with a typical physical resolution of $\sim$ 1.4 kpc. Although EDGE-CALIFA contains galaxies over a wide range in environment, morphological type, and star formation rate, the primary component of the survey is also the MS population. Only a handful of galaxies below the MS with CO detections are included in the sample. Therefore, the ALMaQUEST complements the existing spatially resolved CO datasets by adding a couple of tens more galaxies in the green valley as well as central starbursts \citep{lin20,ell20a}.

The MaNGA datacubes utilized in this work are taken directly from the MaNGA DR15 PIPE3D \citep{san16a,san16b} value-added products \citep{san18}, including the global properties, which are integrated over the MaNGA bundle Field-of-View (FoV), and the spaxel-based measurements of \sigsm~ and emission-line fluxes. By using the Balmer decrement computed at each spaxel and a Milky Way extinction curve with Rv = 3.1 \citep{car89}, we correct the dust extinction of the emission line fluxes. The extinction corrected \ha~flux is converted to the SFR following the
method given by Kennicutt (1998) with a Salpeter IMF. \sigsm~ and \sigsfr~ are then computed using the stellar mass and SFR derived for each spaxel, normalized to the physical area of one spaxel with an inclination correction. We note that for a given spaxel, in principle there are mixed contributions from both the newly formed stars and old stellar populations to the \ha~emission, with the former dominate in the star-forming spaxels whereas the latter dominates the retired spaxels. Without performing a sophisticated decomposition \citep[e.g.,][]{smi21} to distinguish the ionization contributions, the underlying assumption in this work is that the \ha~emission in the star-forming spaxels is purely from active star formation. On the other hand, since the \ha~emission in the retired regions is mostly powered by evolved stars rather than new star formation, the SFR estimated using the \ha~to SFR conversion serves only as an upper limit for retired spaxels \citep{sar10,yan12,sin13,hsi17,bel17,can19,ell21b}.

\subsection{Sample selection and spaxel classficiation \label{sec:sampleselection}}

Traditionally, `green valley' was defined as the sparse region located between the blue cloud and red sequence in the color magnitude diagram \citep[e.g.,][]{wyd07,mar07}. It was later found that depending on the color combination (e.g., UV versus optical), the selected green valley galaxies may or may not well trace the populations that are in the transitional phase from the star-forming to the quiescent populations \citep{sal14, nyi21}. Therefore, alternative methods, such as the location on the SFR vs. \sm~ plane \citep{bel18,jia20,jia21}, sSFR \citep{lin17,sta19}, and D4000 break \citep{ang21}, have been adopted, in addition to the conventional color selections \citep{wyd07,mar07,schi07,sal14,sch14}.  The term `green valley' has now been widely used to refer to the transitioning galaxies in terms of their star formation activities. As our primary goal is to compare the difference between galaxies experiencing quenching with respect to the typical star forming galaxies, we first exclude 12 ALMaQUEST galaxies that were explicitly selected due to their starbursting nature \citep{ell20a} from our analysis. For the remaining 34 galaxies, we group them with a sSFR threshold of $10^{-10.5}$yr$^{-1}$, above and below which correspond to the main sequence (12 galaxies) and green valley (22 galaxies) samples, respectively. As described in \citet{lin20}, all the ALMaQUEST galaxies are detected in CO. In other words, the green valley galaxies used throughout this work are CO-detected galaxies with sSFR below the typical MS galaxies and hence represent transitioning objects.

For the purpose of selecting star-forming spaxels, we first limit our analyses to spaxels having signal to noise (S/N) > 3 for the \ha~and~\hb~ lines and S/N > 2 for the [OIII] and [NII] lines. We then classify each MaNGA spaxel into regions where the dominant ionizing source is star formation, composite, or AGN using the BPT diagnostic based on the [OIII]/\hb~vs. [NII]/\ha~ line ratios \citep{kew01,kew06}. Eq. (3) of \citet{cid10} is then applied to further separate the AGN regions into Seyfert and LI(N)ER regions.  We identify star-forming spaxels to be those satisfying both BPT classified star-forming criteria and the \ha~equivalent width (EW) $>$ 6 \AA~cut. 
The retired spaxels are identified as those having \ha~ EW $<$ 3 \AA~ and S/N $>$ 3 in \ha \footnote{\citet{lin19a} applied a more stringent selection of retired spaxels to be those satisfying both the BPT LI(N)ER criteria and the \ha~EW $<$ 3 \AA~cut in order to ensure that the emissions are primarily photoionized by the old stellar populations. However, this leads to very few retired spaxels in MS galaxies given a small size of the ALMaQUEST MS sample. To increase the number of retired spaxels in MS galaxies, we release the LI(N)ER constraint in this work. Nevertheless, we note that the results and conclusions presented in \S 3 remain very similar even if we follow the selection as done in \citet{lin19a} by imposing the LI(N)ER cut on top of the \ha~EW cut.}. In addition to the aforementioned selection criteria, we also restrict our subsequent analyses to regions having S/N > 3 in the CO line and \sigsm~ > 10$^{7.1}$ (\Msolar~ kpc$^{-2}$).
Even though the PIPE3D stellar products only contain spaxels with continuum S/N >3, our \sigsm~cut helps further remove spaxels with anomalously small values in the stellar mass surface density. These selections lead to the final 6122 (5356) and 108 (1306) star-forming and retired  spaxels in MS (GV) galaxies used in this study, respectively. We note that while the cut in the CO enables the study of the scaling relations in the 3-parameter (\sigsfr, \sigsm, and \sigh2) space, by selection, we will miss retired spaxels that are below the CO detection threshold. We will discuss the potential impact in relevant selections (\S. 3.1.1 and 3.1.3).

\begin{figure}

\includegraphics[angle=0,width=0.5\textwidth]{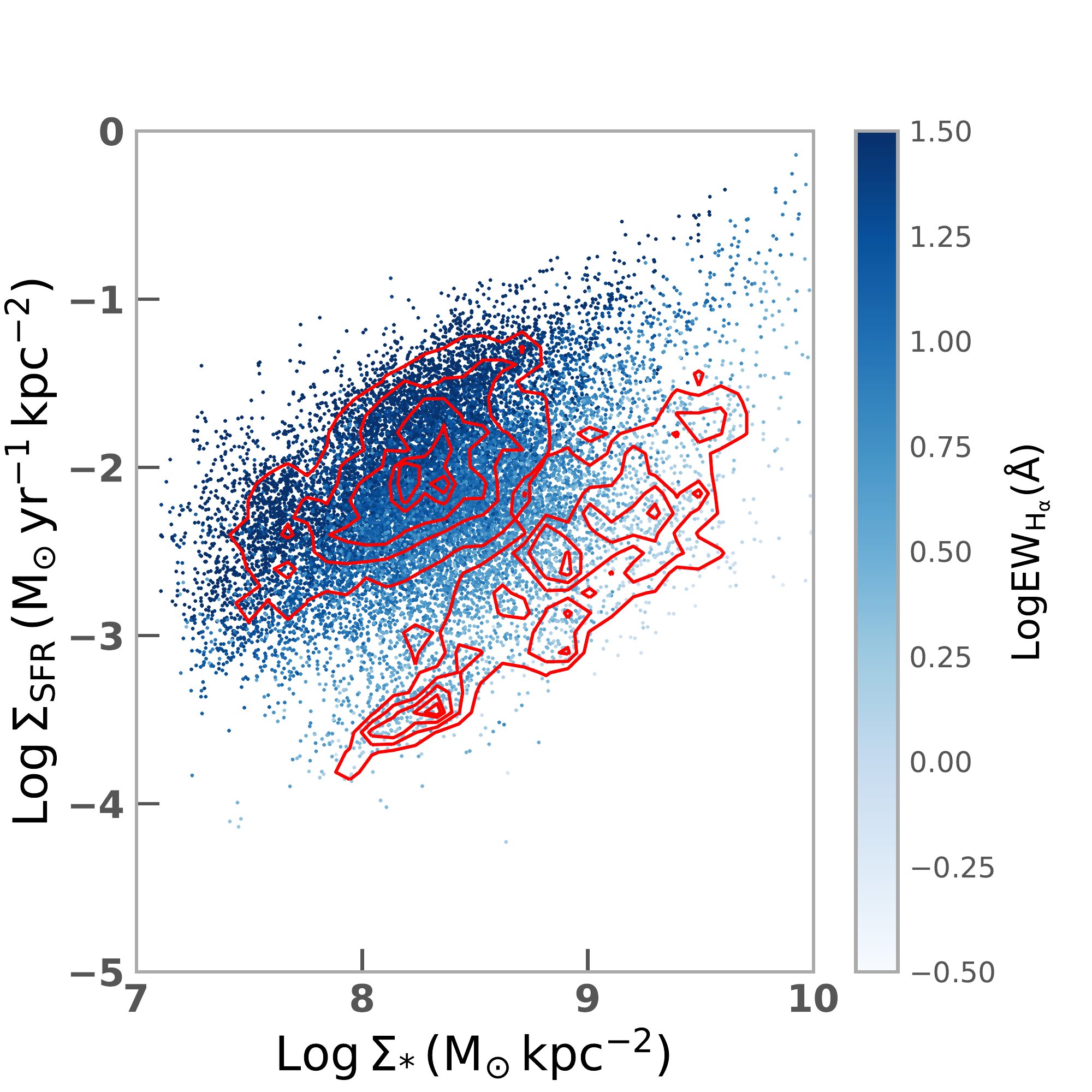}

\caption{Distribution of the 34 ALMaQUEST galaxies on a spaxel-by-spaxel basis in the \sigsfr~and \sigsm~plane, with the colorscale varying according to the \ha~EW in log. The uppwer-left and lower-right red contours represent the distributions for spaxels classified as star-forming and retired regions, respectively. \label{fig:ew}}
\end{figure}

Since the \ha~ EW is the ratio of the line flux (partly if not all due to star formation) to the stellar continuum flux, which approximately scales with stellar mass, it is anticipated that the \ha~ EW is a good proxy of sSFR \citep[e.g., see][]{bel18}. In Figure \ref{fig:ew} we show the spaxel distribution of the 34 galaxies used in this study on the \sigsfr~versus \sigsm~ plane, with the color scaled with the \ha~ EW. It can be seen that both the distributions of sSFR (bearing in mind that sSFR is the ratio between \sigsfr~and \sigsm) and \ha~EW are continuous, decreasing from the upper-left toward the lower-right region on this diagram. In other words, there is a fairly good correspondence between \ha~EW and sSFR.  The two red contours show the distributions of the star-forming spaxels (upper-left contours) and the retired spaxels (lower-right contours) identified using the criteria described. Our selections of star-forming and retire spaxels indeed represent the regions with high and low sSFR, respectively.

\subsection{Bulge and disk decomposition}

\begin{figure*}

\includegraphics[angle=0,width=0.9\textwidth]{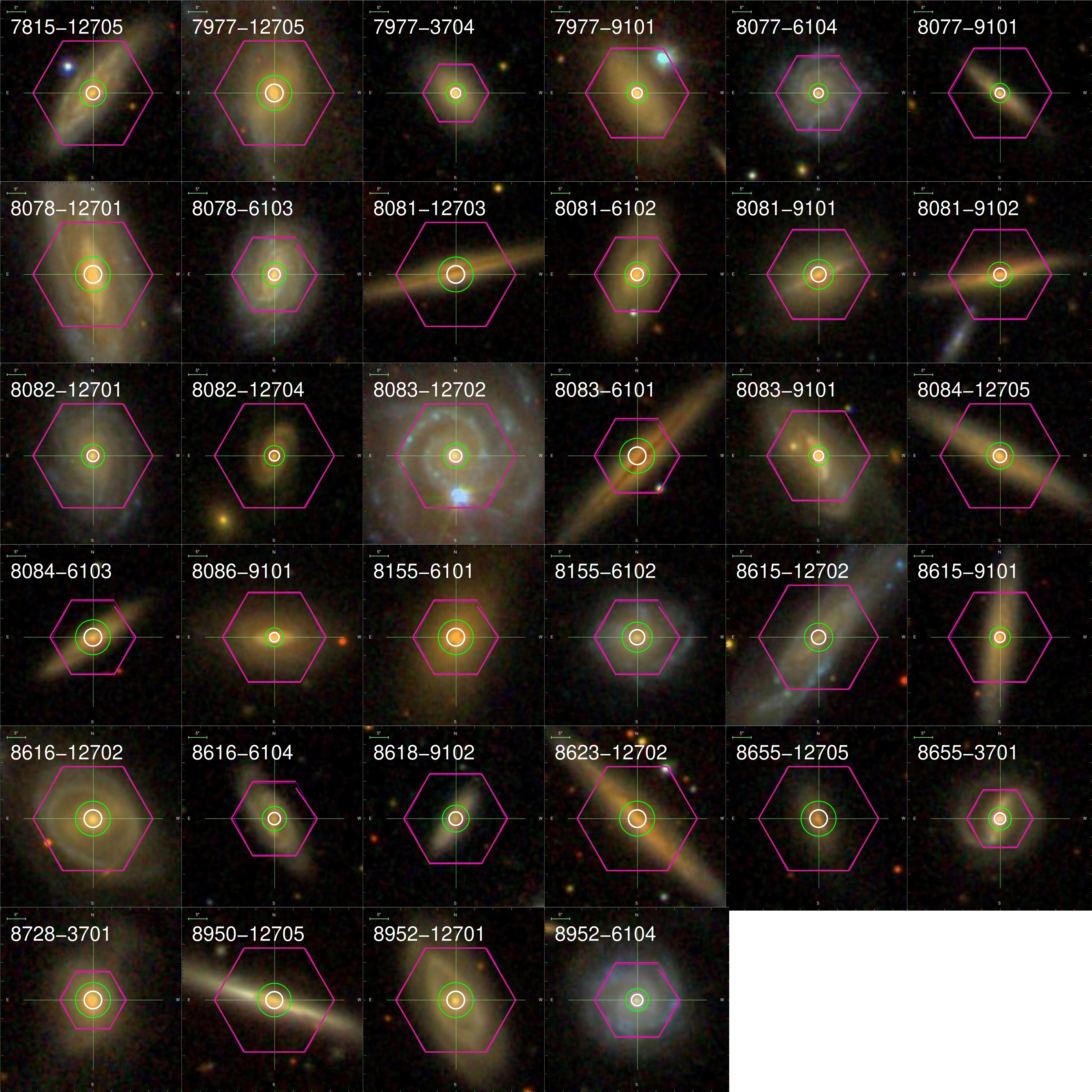}

\caption{The SDSS $gri$ composite images of the 34 ALMaQUEST galaxies used in this work. The magenta hexagons denote the MaNGA footprints. The white circle marks the outter boundary of the areas identified as a bulge region while the green clircle denotes the inner boundary of the disk component. The MaNGA plateifu identifier is shown in white in the upper left corner of each panel. \label{fig:sdss}}
\end{figure*}

As one of our science goals is to compare \fh2~and SFE between the bulge and disk regions, we follow a similar procedure as described in \citet{lin17} to separate the bulge and disk regions for the galaxies used in this work. The two-component fitting is achieved using \texttt{GALFIT} \citet{pen02,pen10} on the SDSS $r-$band images.  We fix the Sersic index to $n = 1$ for the disk component and treat the Sersic index of the bulge component as a free parameter \footnote{On the other hand, the Sersic index was fixed to $n=4$ for the bulge component in the work of \citet{lin17}.}. As the two-component fitting is sensitive to the initial conditions, the determination of the sky level, and the dust obscuration of the images, sometimes the best-fit may yield unrealistic solutions, for example, unphysically large bulge size. As one can see from the SDSS images shown in Figure \ref{fig:sdss}, the majority of our targets are disk-dominated with a small bulge. To ensure that our measurements of the bulge regions are not contaminated by those in the disks, we take a rather conservative approach by imposing an upper limit of the intrinsic bulge size to be 2\arcsec~ in radius \footnote{The typical SDSS seeing (FWHM) is $\sim2.5$ \arcsec~ and therefore a 2 \arcsec~ bulge (in radius) is resolved, but not much.}, corresponding to $\sim$1.2 kpc at the typical redshift ($z \sim 0.03$) of our sample.  Once we obtain the effective radius ($R_e$) of the bulge, we compute the observed effective radius ($R_e^{obs}$) by convolving it with the PSF size of both MaNGA and ALMA beams ($\sim$ 2.5\arcsec). We then define the "bulge" region to be $r <  R_e^{obs}$. The SDSS $gri$ composite images with bulge regions overlaid of the 34 ALMaQUEST galaxies used in this work are shown in Figure \ref{fig:sdss}. We visually checked the derived bulge sizes. We find that the enclosed areas provide a reasonable representation of the bulge regions. Varying the upper limit of the intrinsic bulge size between 1.25\arcsec~ to 4\arcsec~ does not significantly impact any of our conclusions. Although the upper limit of the intrinsic bulge size is checked by eye, in order to mitigate the impact of this choice and/or potential contamination from any overlap region between the bulge and disk, we also define the `disk' region to be $r > $ 2$\times R_e^{obs}$ (denoted by the green circles in Figure \ref{fig:sdss}) when performing subsequent analyses that compare the properties between the bulge and disk regions.

Among the 34 galaxies used in this sample, 7 galaxies potentially host a photometric bar based on visual inspections. Previous works have suggested that the presence of a bar could induce gas inflow, trigger a starburst, and finally lead to the quench of its host \citep[e.g.,][]{alo01,hun08,ell11,car16,ger21}. Therefore, it is possible that the properties of a bar could be distinct from the disk regions and contaminate our results. We find that roughly half of the bars (4/7) in our sample can be classified as retired regions while the other half (3/7) are consistent with composite spaxels. However, as mentioned above, we have imposed a  $r > $ 2$\times R_e^{obs}$ criteria for the selection of disk regions, which alleviates the concern that the contribution from the bars could dominate the results in the disk component.

\section{RESULTS}
\subsection{Scaling relations in star-forming and retired spaxels between MS and GV galaxies\label{sec:scaling}}

Figure \ref{fig:3d} shows the spaxel distributions (individual points) for all spaxels detected in CO and \ha~with S/N > 3 and \sigsm~$>10^{7.1}$ \Msolar~ in the \sigsm--\sigsfr--\sigh2~ 3D space for spaxels in both MS (blue symbols) and GV (green symbols) galaxies. Their projections on the rSFMS, rSK, and rMGMS for MS (blue colors) and GV (green colors) galaxies are also shown as contours. It can be seen that although the spaxels of GV galaxies in general also form sequences in the rSFMS, rSK, and rMGMS 2-D planes, they exhibit a clear offset from the MS galaxies, hinting that there is fundamental difference between MS and GV in terms of their sSFR, SFE, and \fh2. While part of this result is somewhat expected, since the GV galaxies by design are selected to be those with lower global sSFR, the deficit in SFE and \fh2~ for GV galaxies suggests that both effects contribute to the low sSFR of GVs. 

For each galaxy, we compute the fractions of spaxels classified as star-forming or retired. The number of galaxies with a given fraction of star-forming and retired spaxels is shown in the upper and lower panels of Figure \ref{fig:his_ion}, respectively. Compared to MS galaxies (shown as blue bars), GV galaxies (shown as green bars) tend to have a lower fraction of star-forming spaxels and a higher fraction of retired spaxels. One central question is therefore to understand whether the low sSFR of GV galaxies is driven by purely the lower fraction of star-forming spaxels or a global reduction of SFE and \fh2~in all different types of spaxels. In order to address this question, in the following subsections we discuss the three scaling relations for star-forming and retired spaxels, separately for MS and GV galaxies. We fit each of the subsamples using the orthogonal distance regression (ODR) fitting method with a power law parametrized as the following:

\begin{eqnarray}
log_{10} \frac{\Sigma_{SFR}}{\rm M_{\odot}yr^{-1} kpc^{-2}} &=& a*log_{10} \frac{\Sigma_{*}}{10^{8}\rm M_{\odot} kpc^{-2}} + b\\
log_{10} \frac{\Sigma_{SFR}}{\rm M_{\odot}yr^{-1} kpc^{-2}} &=& a*log_{10} \frac{\Sigma_{H2}}{10^{7}\rm M_{\odot} kpc^{-2}} + b\\
log_{10} \frac{\Sigma_{H2}}{\rm M_{\odot}kpc^{-2}} &=& a*log_{10} \frac{\Sigma_{*}}{10^{8}\rm M_{\odot} kpc^{-2}} + b
\end{eqnarray}

where $a$ and $b$ denote the slope and the normalization of the scaling relations in log, respectively. The obtained best-fit parameters ($a$ and $b$) for various subsamples presented in \S \ref{sec:rSFMS} -- \S \ref{sec:rMGMS} are listed in Table \ref{tab:fit}. It is worth noting that the measurements of each spaxel are not fully independent, though, since the size of spaxel (0.5$\arcsec$) is oversampled by a factor of $\sim 20$ with respect to the PSF size ($\sim$2.5$\arcsec$). However, the types of the analyses presented in this work is found to be robust against the binning resolution \citep[e.g., see][]{ell21a}. On the other hand, the uncertainty of the fits could be underestimated due to the oversampling. To account for this effect, we take a rather conservative approach. We multiply the derived uncertainty of the fits by a factor of square root of the oversampled factor, assuming that the oversampled datapoints are highly correlated. The reported uncertainties in subsequent sections and Table \ref{tab:fit} therefore serve as an conservative estimate.

\subsubsection{The resolved star forming main sequence relation (rSFMS)\label{sec:rSFMS}}
We first compare the rSFMS between star-forming and retired spaxels for MS and GV galaxies separately. Figure \ref{fig:rSFMS} shows the rSFMS of star-forming spaxels (left panels) and retired spaxels (right panels) for MS (top panels) and GV (bottom panels) galaxies. The blue dashed line represents the rSFMS derived from star-forming spaxels in main-sequence galaxies (i.e. a fit to the points in the top left panel) and is the same reference line shown in all panels to guide the eyes. The derived slope for the star-forming spaxels of the MS galaxies is 1.11, in good agreement with previous results using the ALMaQUEST sample, despite that the sample selection and the S/N ratio cuts in the final spaxels differ slightly in detail \citep{lin19b,ell20a}. Our result is also in broad agreement with the slopes found in other studies, ranging from 0.6 to 1.3 \citep{san13,can16,gon16,hsi17,err19,can19,mor20,eni20,san21}. 

The solid lines are the best fit of the data points in a given panel (blue: star-forming spaxels; red: retired spaxels). The retired spaxels in either MS or GV galaxies are both found to lie below the MS rSFMS curve, offset by nearly one order of magnitude. One thing to note is that as we mentioned earlier, the MS galaxies are dominated by the star-forming spaxels and hence the number of retired spaxels used here is very limited. A larger sample would be beneficial to better characterize various properties of the retire spaxels in the MS galaxies.

The slope of the retired sequence is also found to be close to unity, consistent with other resolved studies \citep{hsi17,can19}.  The lower values of \sigsfr~ (or \ha~flux) at a given \sigsm~ in retired spaxels are previously known as the `LI(N)ER' or retired sequence \citep{hsi17,can19}, in which the \ha~emissions are attributed to contributions primarily from evolved stars rather than new star formation \citep{sar10,yan12,sin13,bel17}. As we mentioned earlier, the \ha-based SFR for retired spaxels is therefore considered to be an upper limit \citep[e.g.,][]{can19}. As a result, the actual difference in \sigsfr~between star-forming and retired spaxels at a given \sigsm~could be even greater than what we see here. Also as mentioned in \S \ref{sec:sampleselection}, our selection of retired spaxels does not contain regions without CO detections, i.e., regions not forming stars at all due to the lack of cold molecular gas. If included, they would further enlarge the difference in \sigsfr~ between star-forming and retired spaxels at a given \sm.

Another crucial feature revealed in Figure \ref{fig:rSFMS} is that the \sigsfr~ of star-forming spaxels in GV galaxies is also found to be lower than that in MS galaxies at a given \sigsm, suggesting that the star formation in GV galaxies is suppressed even in regions classified as star-forming. The difference in the derived slopes and the normalizations between MS and GV are greater than 6 sigma (see Table \ref{tab:fit}).  Our result therefore indicates that the depletion in the sSFR of GV galaxies may appear to be a global phenomenon within the galaxies, not restricted to certain areas. This is consistent with some of the recent complementary works that studied the SFR radial profile of GV galaxies and found a global reduction with respect to MS galaxies at all radii \citep[e.g.,][]{bel18,ell18,bro20}. 

\subsubsection{The resolved Schmidt-Kennicutt relation (rSK)\label{sec:rSK}}
Figure \ref{fig:rSK} compares the rSK relation of star-forming and retired spaxels for both MS and GV galaxies. It is evident that the retired spaxels of either MS or GV galaxies tend to lie below the rSK relation formed by the MS star-forming spaxels, in other words, at lower SFE. When comparing the star-forming spaxels between MS and GV galaxies (the upper left and lower left panel, respectively), we find that the GV star-forming spaxels also deviate from the MS rSK curve towards a lower \sigsfr~at a given \sigh2. 
The best-fit of the slope ($a$ = 0.81$\pm$0.05) and the normalization ($b$ = -2.13$\pm$0.02) of star-forming spaxels in GV galaxies is found to be at least 4 sigma away from that ($a$ = 1.02$\pm$0.03;  $b$ = -1.94$\pm$0.01) in MS galaxies.
In other words, the efficiency of converting the molecular gas into stars in the GV galaxies  is suppressed not only in retired but also in star-forming spaxels. This is in good agreement with a previous finding that the deviation of rSK from the ensemble relation strongly correlates with the global sSFR \citep{ell21a}.

\subsubsection{The resolved molecular gas main sequence relation (rMGMS)\label{sec:rMGMS}}
The relationship between \sigh2~and \sigsm~at (sub)kpc scales has been investigated recently to gain insight into the role of molecular gas abundance in shaping galaxy properties \citep[e.g.,][]{won13,lin19b,bar20,mor20,ell21a,san21,pes21}. It was found that these two quantities for star-forming regions in nearby galaxies form a very tight relation, dubbed the rMGMS \citep[e.g.,][]{lin19b}.
Recently, \citet{ell21b} extended the study of rMGMS to retired spaxels by using 8 ALMaQUEST galaxies and showed that \fh2~of retired spaxels is systematically lower by a factor of $\sim5$ than that of star-forming spaxels. Similarly, a depletion of \fh2~is also found in central AGN regions relative to the star-forming regions in EDGE-CALIFA \citep{ell21c}. Here we take a step further to investigate the variation of molecular gas fraction by separating the galaxies into MS and GV categories with a larger ALMaQUEST sample. Figure \ref{fig:rMGMS} is analogous to Figures \ref{fig:rSFMS} and \ref{fig:rSK} but for the rMGMS relation. Similarly, the retired spaxels of either MS or GV galaxies occupy the regions below the reference relation formed by the MS star-forming spaxels, hinting at a lower gas fraction in the retired regions, as previously reported by \citet{ell21b}. As mentioned in \S \ref{sec:sampleselection}, our retired spaxels do not include those falling below the CO S/N cut, which would lie further below from the rMGMS reported here, in other words, even lower gas fraction.

Meanwhile, the star-forming spaxels of GV galaxies are also found to lie below the reference line (i.e., the rMGMS of star-forming spaxels in MS galaxies). While the slopes between the GV star-forming spaxels and the MS star-forming spaxels are both close to unity (0.97$\pm$0.04 vs. 1.06$\pm$0.04), there is 8-sigma difference in the normalization (6.66$\pm$0.01 vs. 6.82$\pm$0.02). Our finding is consistent with a complementary analysis performed by \citet{ell21a}, who found a dependence of the offset from the median rMGMS relation on the global sSFR.

\subsubsection{Distributions of sSFR, SFE, and \fh2}
Having inspected the distributions of star-forming and retired spaxels for both MS and GV galaxies on the rSFMS, rSK, and rMGMS planes, we then compare the ratios of the parameters of these relations, i.e., the sSFR, SFE, and \fh2, in star-forming and retired spaxels to better quantify the differences. Figure \ref{fig:his_gas} shows the distribution of spaxel-based sSFR (left panels), SFE (middle panels), and \fh2 (right panels) for star-forming (blue) and retired (red) regions. The upper and lower three panels are spaxels belonging to the MS and GV galaxies, respectively. The purple dotted line represents the median value for the star-forming spaxels in MS galaxies as a reference line and is the same between the corresponding upper and lower panels. The median values of the associated histograms, $m$, are reported in each panel. In both GV and MS galaxies, the retired spaxels show lower values than those of star-forming spaxels by $\sim$ 1.1, 0.6, and 0.5 dex in sSFR, SFE, \fh2, respectively. Bearing in mind that the SFE and sSFR estimates in retired regions are conservative upper limits, the actual offsets could be even larger. Our results therefore suggest that the deficit in the sSFR of retired regions relative to the star-forming regions can be attributed to not only the reduction in the gas fraction as reported by \citet{ell21b} but also the reduced SFE.

An important feature clearly seen from Figure \ref{fig:his_gas} is that the star-forming spaxels in GV galaxies also exhibit lower sSFR, SFE, and \fh2~compared to the star-forming spaxels in MS galaxies by 0.36, 0.14, and 0.21 dex, respectively. We perform a Kolmogorov-Smirnov test (KS test) to further examine whether the distributions in the above quantities of star-forming spaxels in the MS galaxies relative to that in the GV population are significant or not. The resulted $p$-value are 0.0, 3.94e-177, and 1.87e-273 for sSFR, SFE, and \fh2~, respectively. The tiny values of $p$-value rules out the null hypothesis that the two samples (star-forming spaxels of MS galaxies vs. star-forming spaxels of GV galaxies) can be drawn from the same population. In other words, the gas properties and hereby the star formation rates in GVs are different from MS galaxies not only because there are fractionally more quenched areas in GVs but also because the gas contents are also different in the star-forming regions between GV and MS populations.

\subsection{SFE and \fh2~in the disk and bulge regions of GV galaxies}

Having studied the scaling relations and the offset in sSFR/SFE/\fh2~ of green valley galaxies relative to the main sequence galaxies, we now turn to the investigation of the dependence of SFE and \fh2~on the location within galaxies. 
For each of the ALMaQUEST MS and GV galaxies, we measure the median values of spaxel-based sSFR, SFE and \fh2~in the bulge and disk regions separately. Since the `bulge' region defined in this analysis refers to the inner region enclosed by a certain radius (either the effective radius of the bulge or a given upper limit; see \S. 2), it is important to keep in mind that we may not be able to remove the disk contribution in the inner regions. Another thing to note is that although the ALMaQUEST target selection does not impose a morphology criterion, the majority of our sample tends to have  a small bulge. Therefore, we are not able to investigate in detail how the star formation and/or gas properties of galaxies depend on the presence of bulges, as explored in \citet{koy19}. With these caveats in mind, we only focus on the comparison of the sSFR, SFE and \fh2~between the MS and GV populations for a given structural component, instead of attempting to discuss the impact of the bulge on these quantities. 

\begin{figure*}
\centering
\gridline{\fig{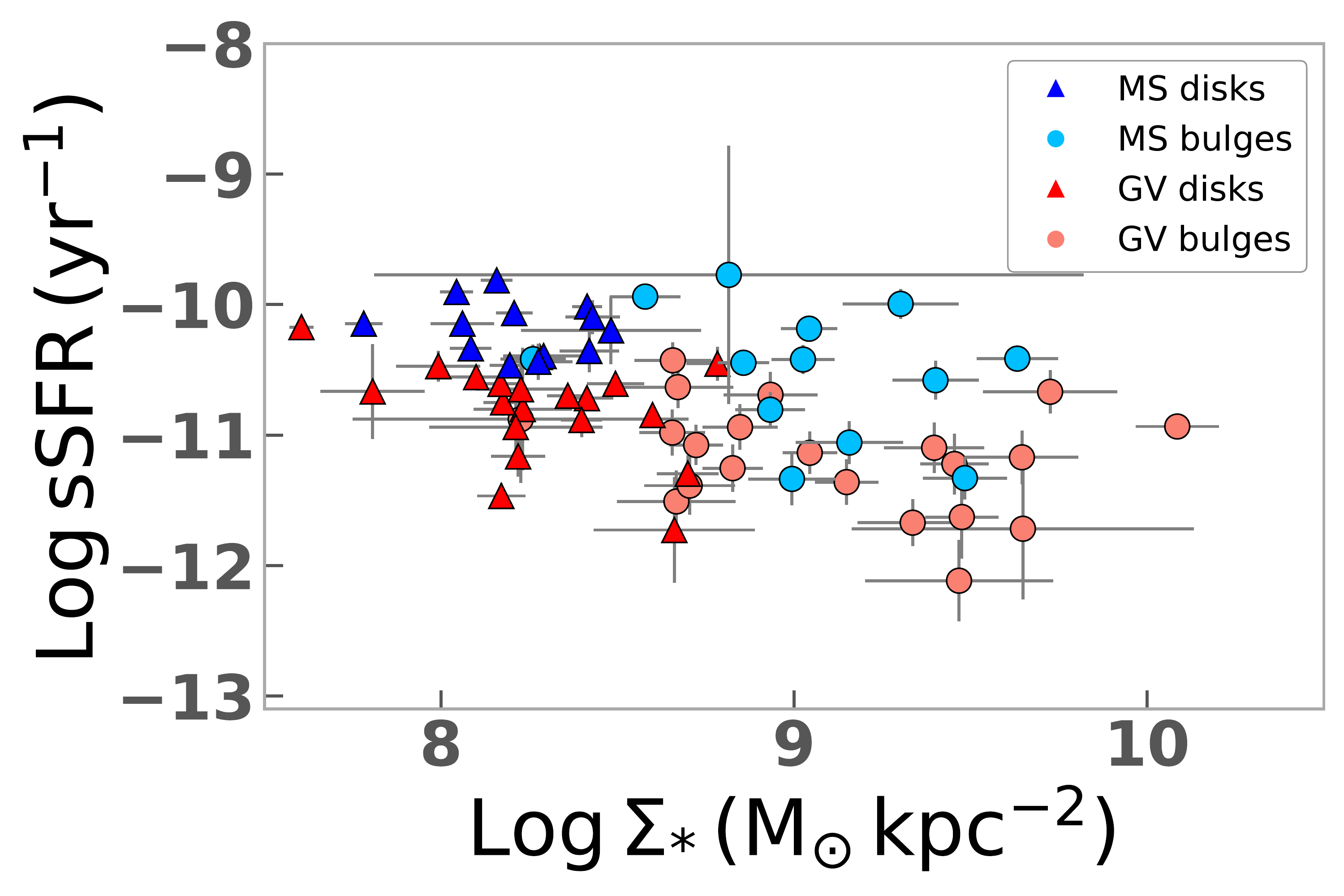}{0.32\textwidth}{}
          \fig{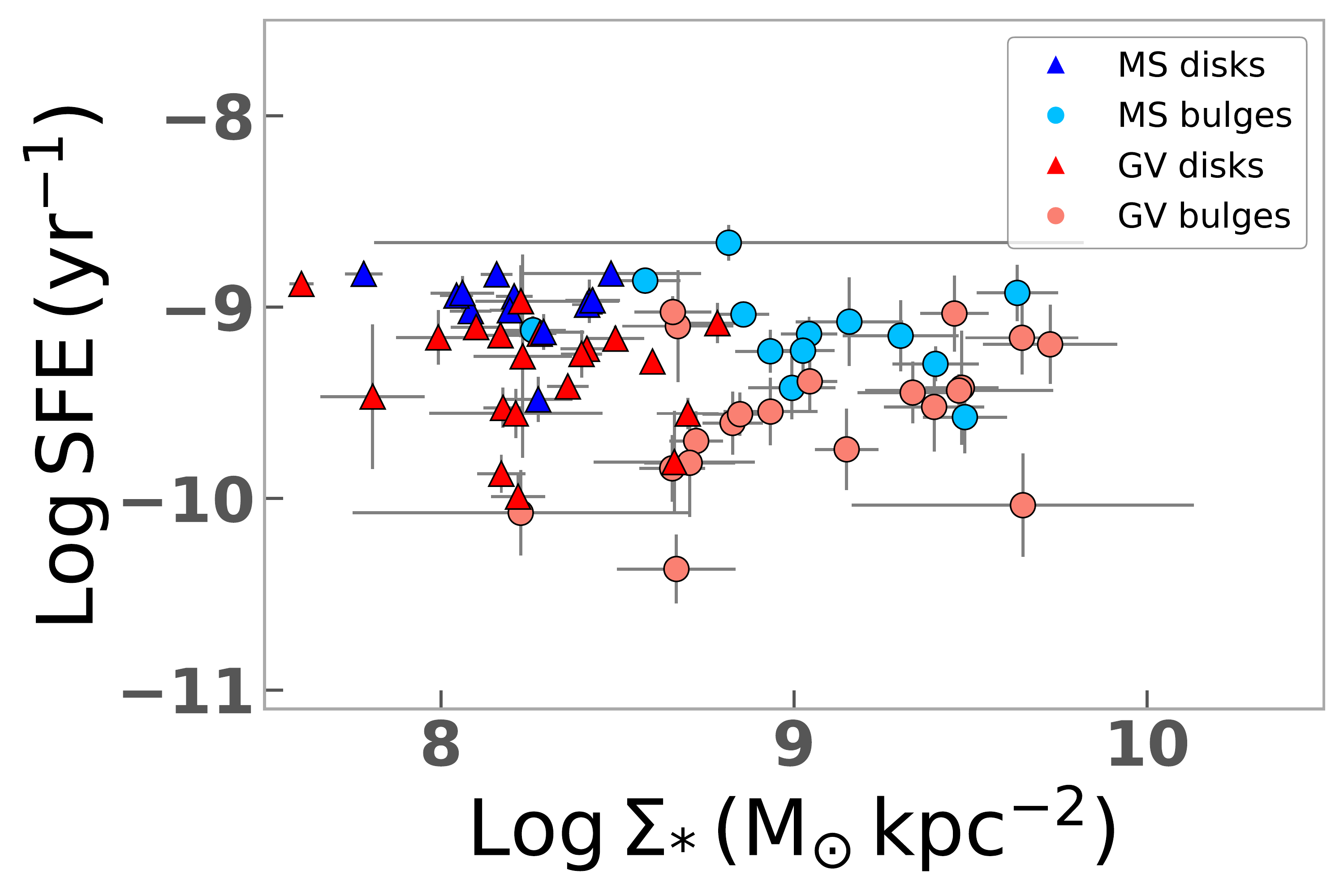}{0.32\textwidth}{}
          \fig{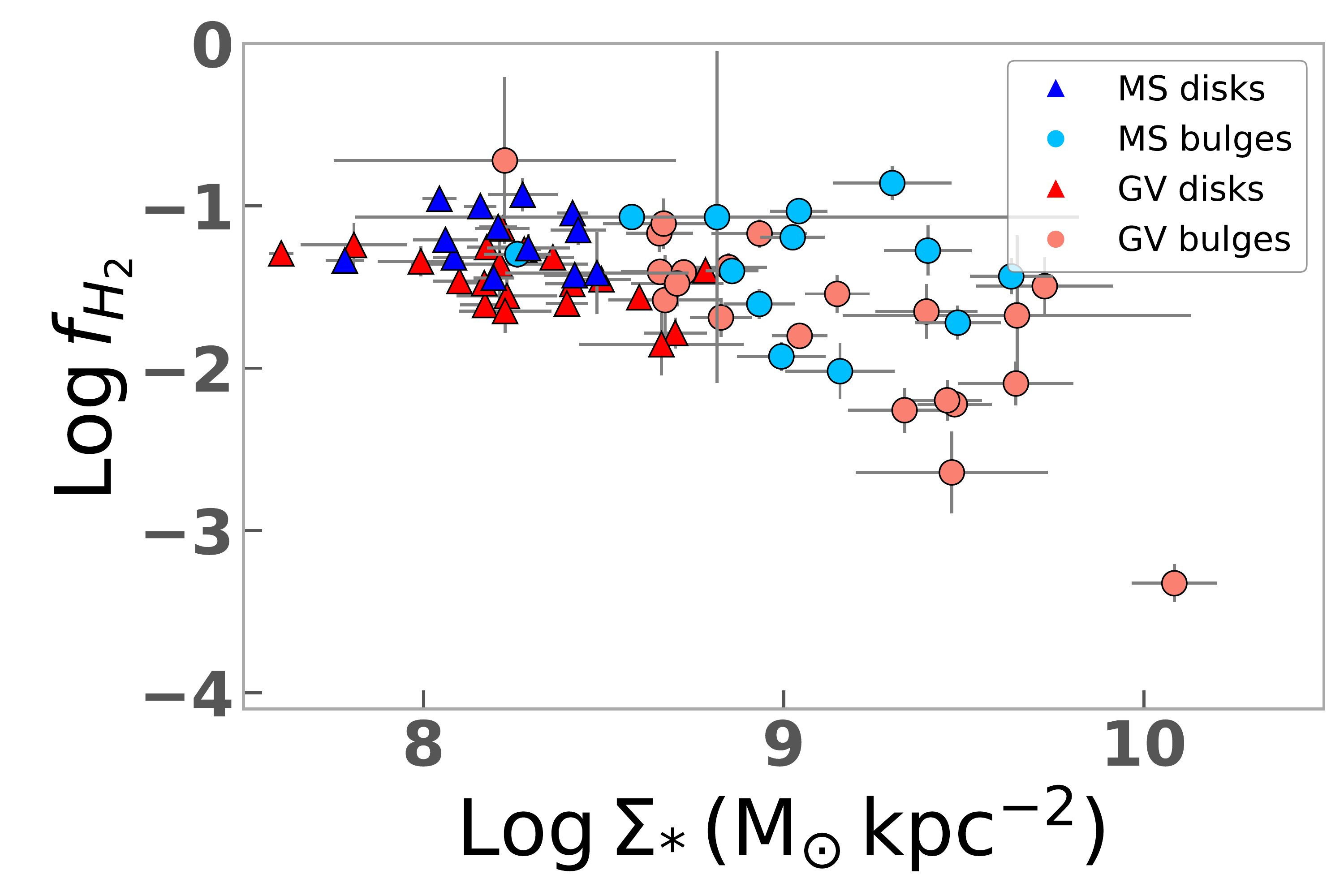}{0.32\textwidth}{}}
          
\caption{The sSFR (left panel), SFE (middle panel), and \fh2~(right panel) as a function of \sigsm. In all the panels, bulges and disks are shown as light circles and dark triangles, respectively. The MS galaxies are shown in blue and GV galaxies are shown in red. \label{fig:bd}}
\end{figure*}

Figure \ref{fig:bd} compares the median spaxel values of sSFR (left panel), SFE (middle panel), and \fh2~(right panel) of the bulge (light circles) and disk (dark triangles) regions between MS (blue colors) and GV (red colors) galaxies. As bulges tend to have larger values of \sigsm~compared to the disks and therefore may occupy distinct regions in terms of the \sigsm~range, we plot sSFR, SFE, and \fh2~ as a function of \sigsm~in Figure \ref{fig:bd} for the clarity of the data points. Visually it is clear from the left panel that there is an apparent sSFR offset between the MS and GV galaxies, either in the disk or in the bulge regions. In other words, while the averaged lower global sSFR of GV compared to MS galaxies is by design, what we show here is that the depleted sSFR is present in both the bulge and disk regions. To quantify the amount of sSFR depletion, we bootstrap the measurement of the sSFR separately for the bulge and disk components of MS and GV galaxies, each with 10000 iterations. We find that the sSFR in GV galaxies is reduced by 0.61 $\pm$0.10 dex and 0.65$\pm$0.16 dex in the disk and bulge regions, respectively, compared to that in MS galaxies. While the relative offset seems comparable between the bulges and disks, it is worth noting that in both MS and GV galaxies, the absolute sSFR values in the bulges are smaller than those in the corresponding disks. In other words, both MS and GV galaxies possess older stellar populations in the central regions compared to outer part of galaxies. 

The middle panel of Figure \ref{fig:bd} shows that the values of SFE for GV galaxies are systematically lower than those of the MS galaxies, either in the disk or bulge regions. By repeating the bootstrapping, we find that the median values of log$_{10}$(SFE/yr$^{-1}$) in the disks are -8.99$\pm$0.05 and -9.34$\pm$0.06 for MS and GV, respectively. On the other hand, the medians of log$_{10}$(SFE/yr$^{-1}$) in the bulge components are found to be -9.13$\pm$0.06 and -9.55$\pm$0.08 for MS and GV, respectively. In other words, GV galaxies show a reduction of SFE in the disk and bulge with respect to the MS galaxies by 0.35$\pm$0.08 dex and  0.42$\pm$0.10 dex, respectively. This trend suggests that the depleted SFE we see in \S \ref{sec:scaling} is a global feature, persisting from the inner to the outer regions. 

In the right panel of Figure \ref{fig:bd}, we plot the median \fh2~ of the bulge and disk regions, separately for MS and GV galaxies. Again we determine the median values of \fh2~through the bootstrapping analysis. The medians (in log) in the disks are found to be -1.20$\pm$0.05 and -1.45$\pm$0.04 for MS and GV, respectively, whereas in the bulges the median values of \fh2~(in log) are determined to be -1.38$\pm$0.09 and -1.71$\pm$0.12 for MS and GV, respectively. This leads to a \fh2~depletion in GV relative to MS by a factor of 0.25$\pm$0.06 dex and 0.34$\pm$0.16 in the disk and bulges, respectively. The reduced \fh2~in the bulge regions of GV relative to MS, however, is only significant at a 2-sigma level. We then additionally run a KS test comparing the \fh2~distributions of the bulge component between MS and GV galaxies and obtain a $p$-value of 0.16. The non-negligible $p$-value, together with the 2-sigma significance in the \fh2~offset between MS and GV indicates that the difference in \fh2~of bulge is only marginal, likely due to the small statistics.

In summary, we find a similar trend as reported in \citet{lin17} and \citet{bro20} that statistically, GV galaxies exhibit lower \fh2~and SFE across the entire galaxy compared to that of MS galaxies, and both low \fh2~and SFE contribute to the sSFR suppression in either the disk or the bulge region of GV galaxies. The reduction of \fh2~in the GV bulges, however, awaits confirmation with a larger sample in the future.

The correlation between SFE and \fh2~has been studied not only using global quantities \citep{pan18b,dou21} but also at kpc scales \citep{bro20,ell20a}. Next we investigate whether the two quantities, \fh2~and SFE, are correlated with each other or not in our sample. In Figure \ref{fig:bd_sfefgas} we present the relation between \fh2~and SFE on a galaxy-by-galaxy basis, separately for MS (blue symbols) and GV (red symbols). Triangles and  circles denote the disk and bulge regions, respectively. The Pearson correlation coefficient ($r_{s}$) and Spearman correlation coefficient ($\rho$) are found to be 0.10 and 0.24 between SFE and \fh2, respectively, indicating that there is no correlation between the strength of these two quantities.
We also note that there are several cases that show high \fh2~but low SFE, representing galaxies where the molecular gas is still present but unable to form stars efficiently. Overall the large scatter in the SFE vs. \fh2~relation suggests that the relative contributions between the SFE and \fh2~in lowering the sSFR in GV galaxies vary from case to case.

\section{SUMMARY AND DISCUSSION  \label{sec:summary}}

In this work, we present the kpc-scale scaling relations of the rSFMS, rSK, and rMGMS for 12 main sequence (MS) and 22 green valley (GV) galaxies drawn from the ALMaQUEST survey in order to investigate the distributions of sSFR, SFE, and \fh2. For each of the two categories of galaxies (i.e., MS and GV), we identify in total $\sim11,500$ star-forming and $\sim1,400$ retired regions on spaxel-by-spaxel basis. Our main findings are:\\

\begin{figure*}

\begin{interactive}{animation}{sfgv_3d.mp4}

\includegraphics[angle=0,width=0.9\textwidth]{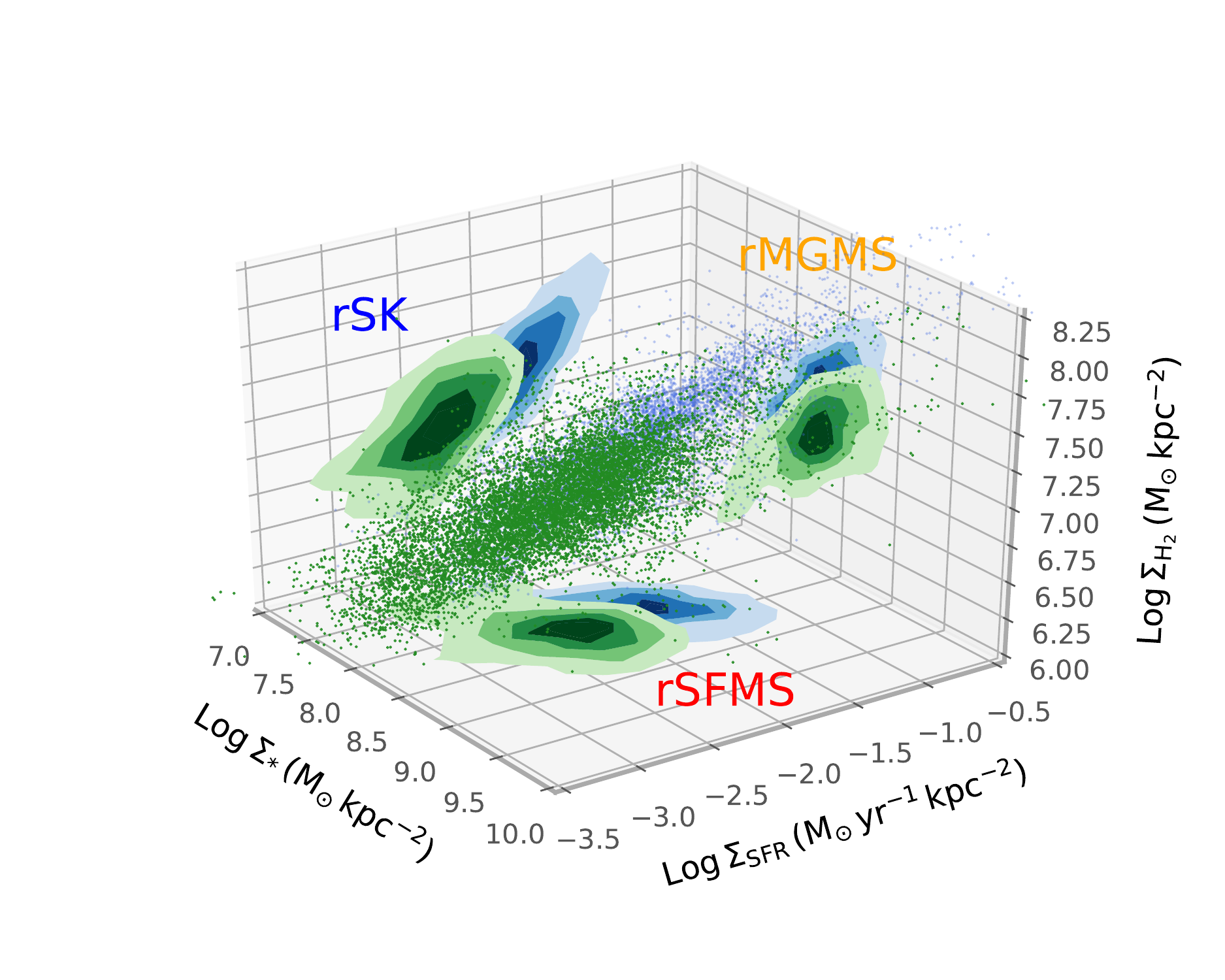}

\end{interactive}
\caption{The 3D distribution between \sigsm, \sigh2, and \sigsfr~displayed for spaxels in 12 MS (blue dots) and 22 GV (green dots) galaxies in ALMaQUEST. The contours (blue for spaxels in MS and green for spaxels in GV galaxies) show the projections on the 2D planes, representing the rSFMS, rSK, and rMGMS scaling relations, respectively. The contour levels correspond to 20\%, 40\%, 60\%, 80\%, and 90\% of the density peaks. An animated version of this figure is available online. The animation shows one rotation about the Z axis. The realtime duration is 18 seconds.
 \label{fig:3d}}
\end{figure*}

\begin{figure}
\centering
\includegraphics[angle=0,width=0.5\textwidth]{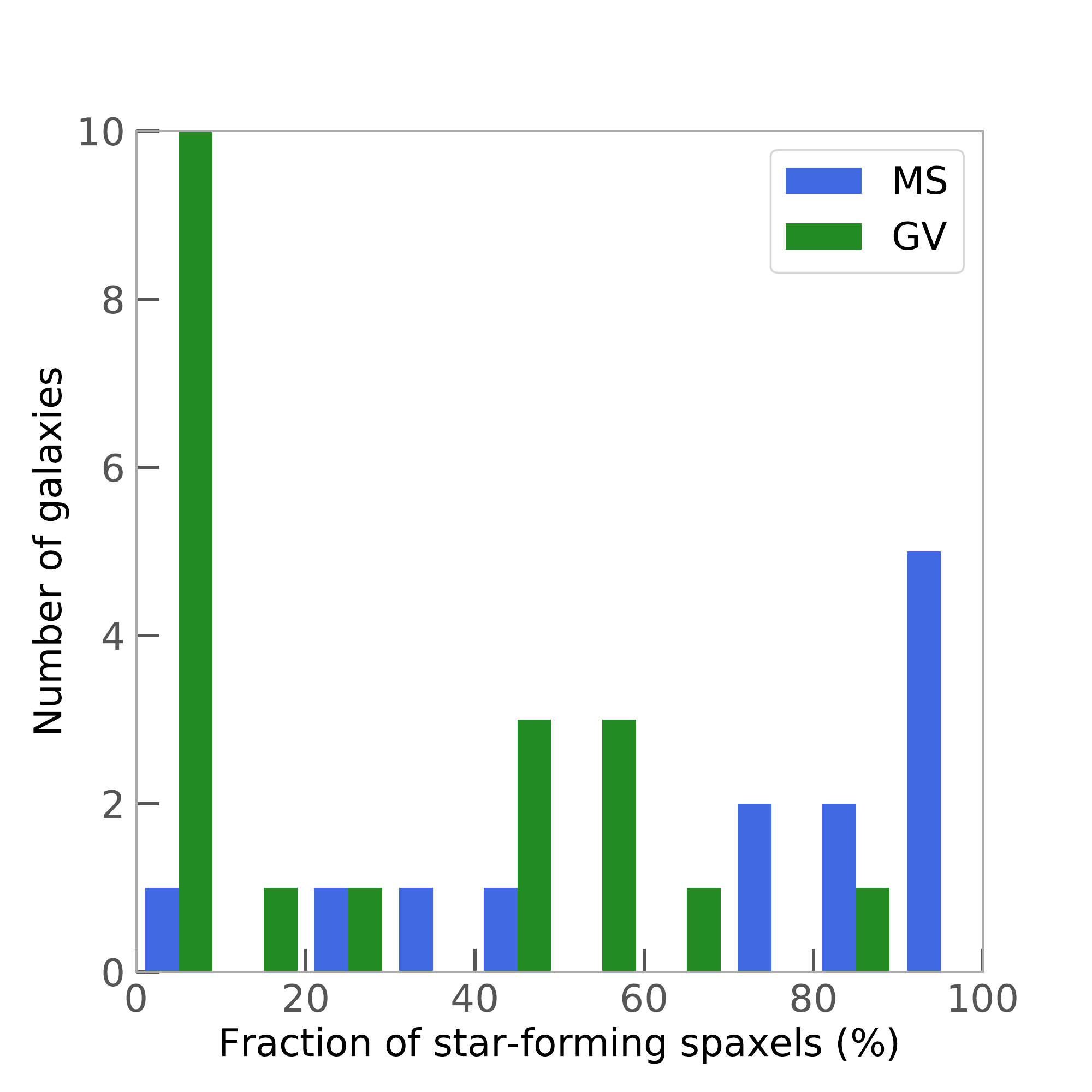}
\includegraphics[angle=0,width=0.5\textwidth]{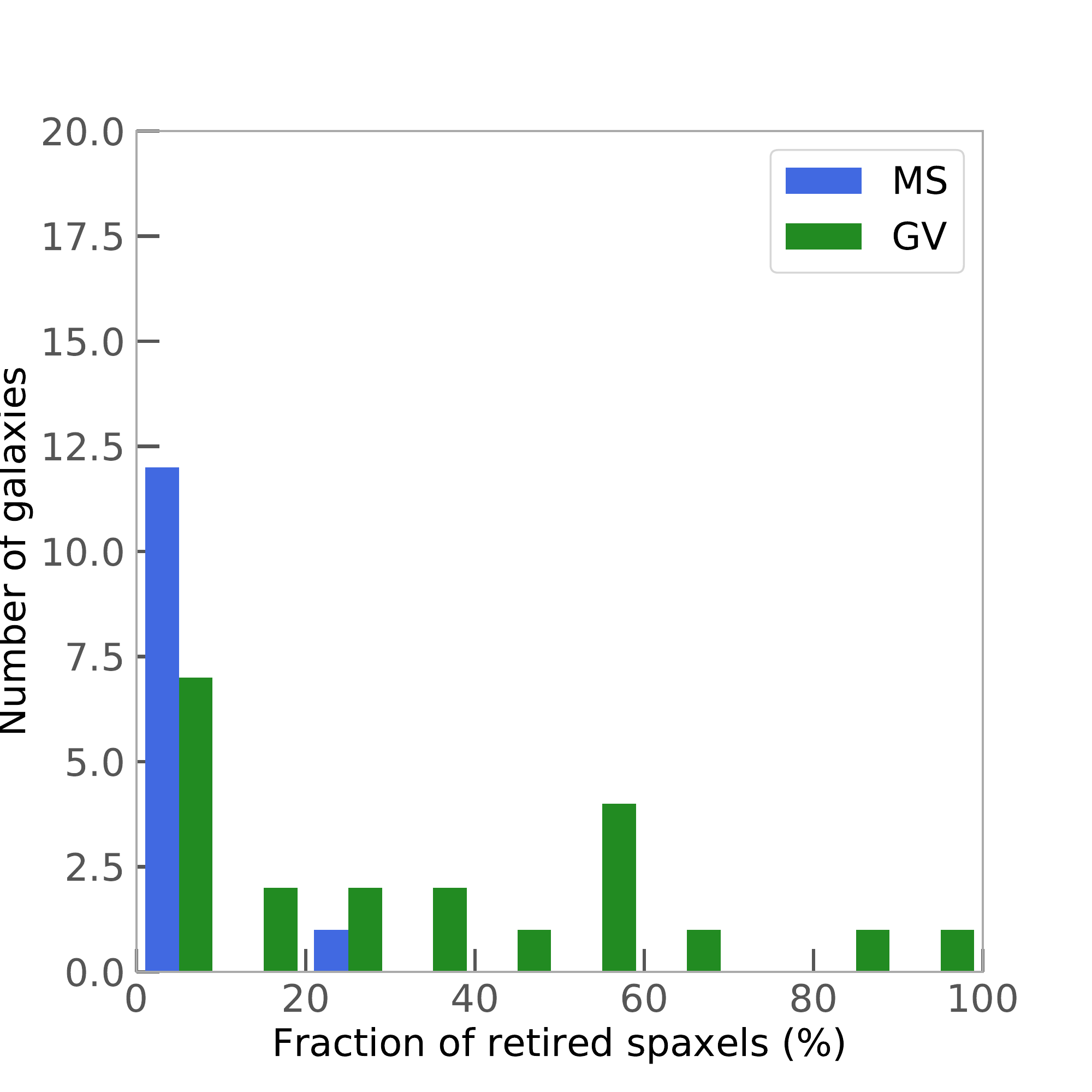}

\caption{Number of galaxies as a function of the percentage of star-forming (upper panel) and retired (lower panel) spaxels. MS galaxies are shown in blue whereas GV galaxies are shown in green. \label{fig:his_ion}}
\end{figure}

1. The distribution of spaxels of GV galaxies in the 3-dimensional space formed by the \sigsfr, \sigsm, and \sigh2, is offset from that in MS galaxies (Figure \ref{fig:3d}). The fraction of star-forming (retired) spaxels decreases (increases) from MS to GV galaxies (Figure \ref{fig:his_ion}).

\begin{figure*}
\gridline{\fig{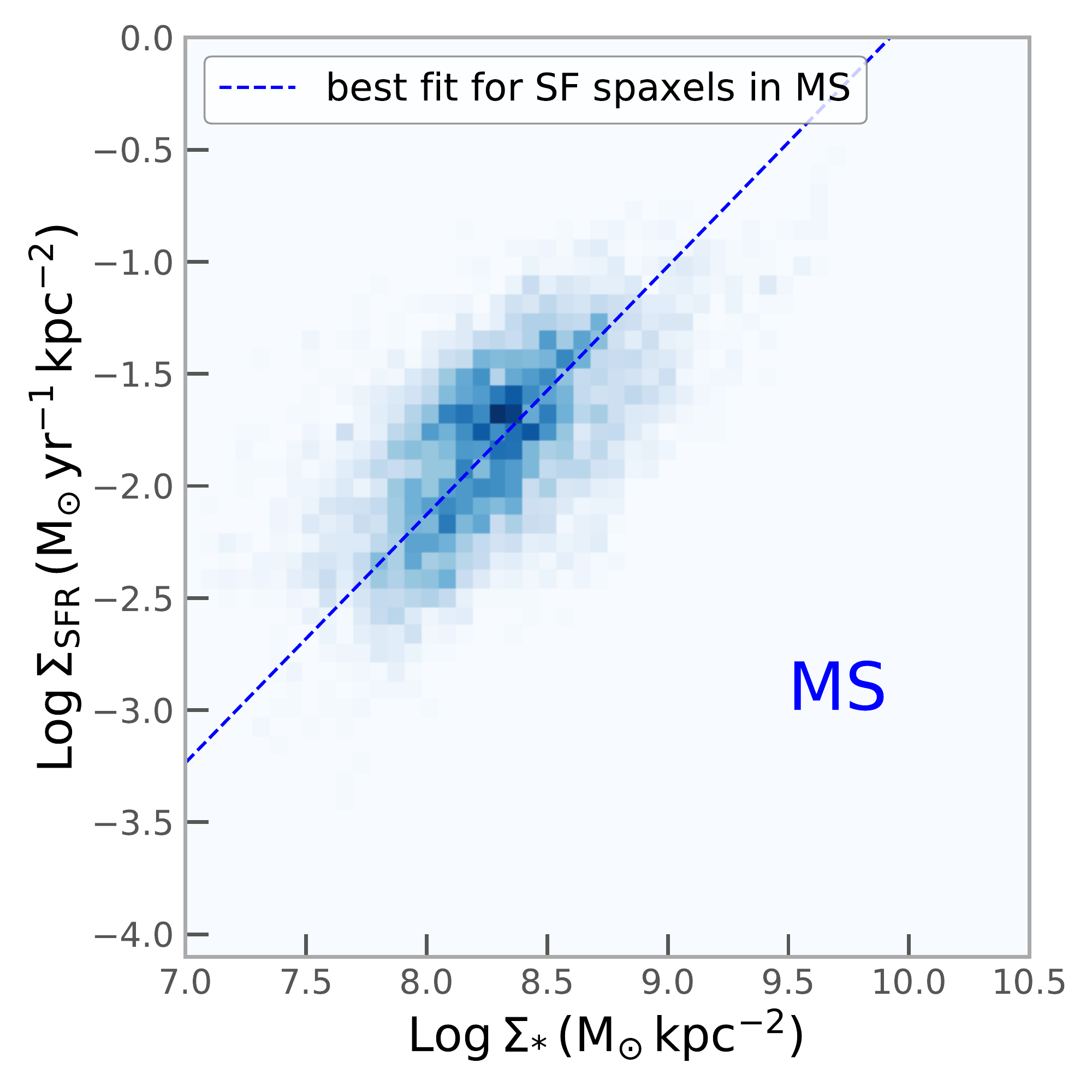}{0.45\textwidth}{(a) SF spaxels in MS galaxies}
          \fig{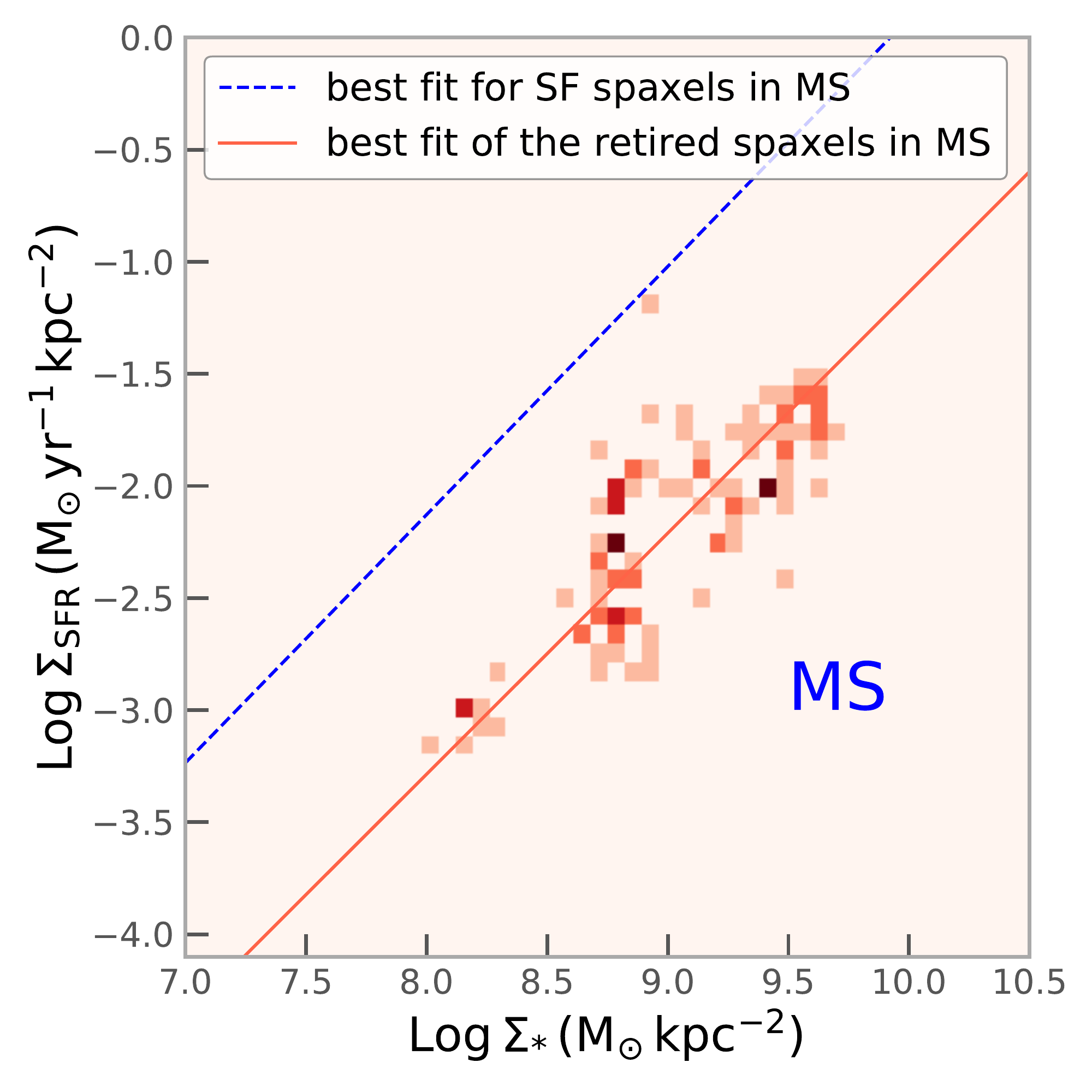}{0.45\textwidth}{(b) Retired spaxels in MS galaxies}}
\gridline{\fig{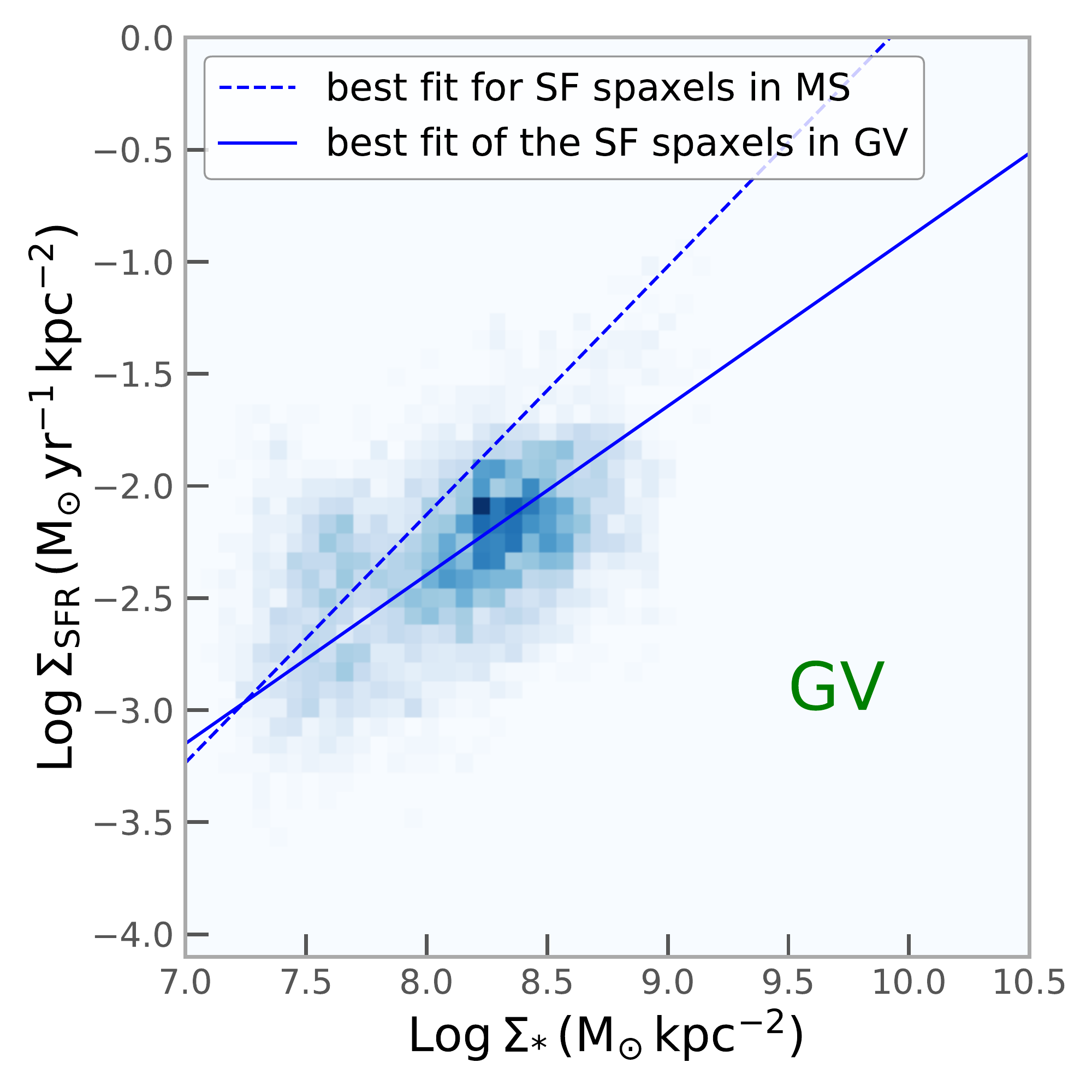}{0.45\textwidth}{(c) SF spaxels in GV galaxies}
		  \fig{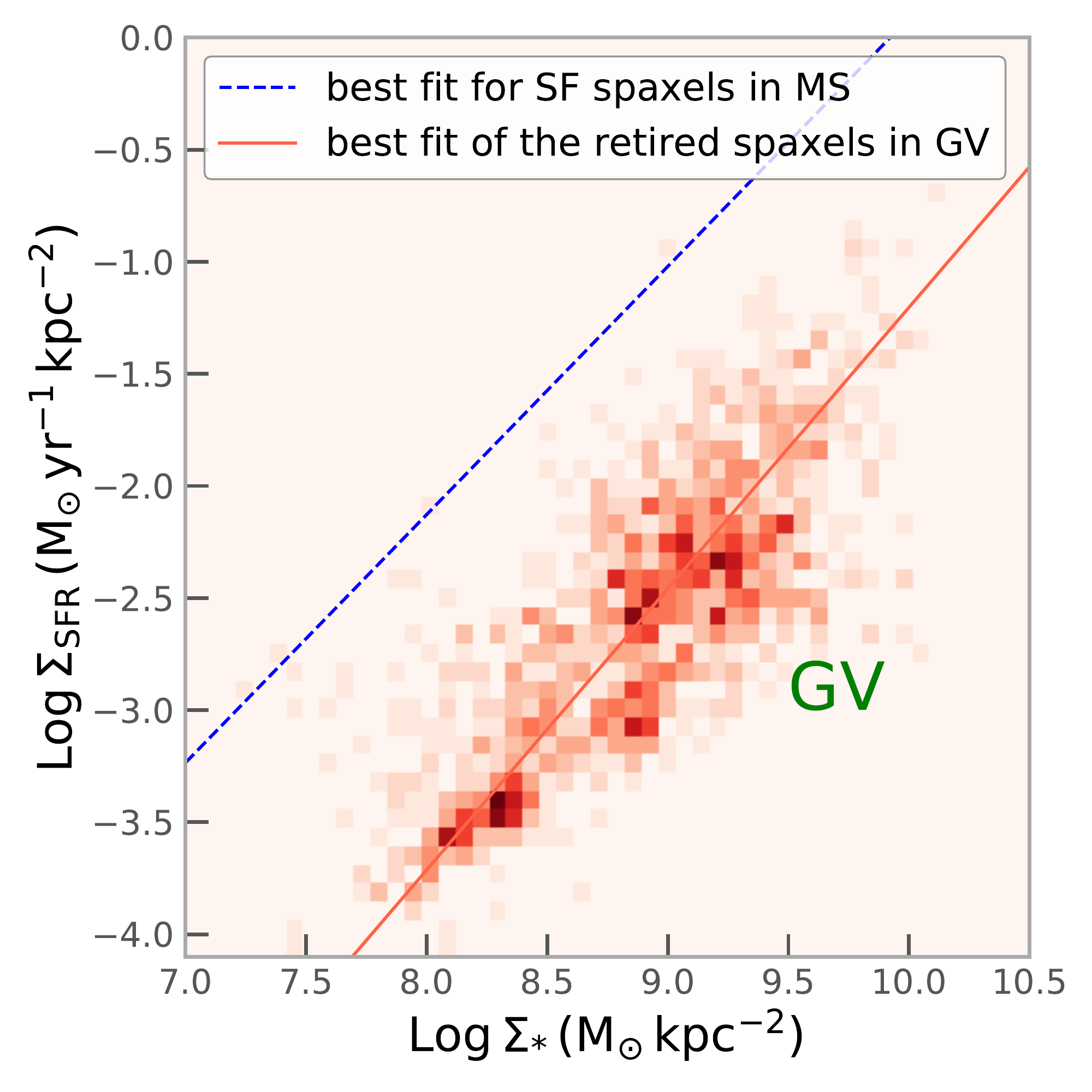}{0.45\textwidth}{(d) Retired spaxels in GV galaxies}}

\caption{Number densities of star-forming (left panels; blue scales) and retired (right panels; red scales) spaxels on the rSFMS plane for MS (top panels) and GV (bottom panels) galaxies. The solid lines represent the best fit of the data points (blue for star-forming spaxels and red for retired spaxels) shown in each panel. The best-fit parameters are listed in Table \ref{tab:fit}. The blue dashed lines are the same reference lines in each panel, corresponding to the best fit of the rSFMS derived for the star-froming spaxels in MS galaxies (i.e. a fit to the spaxels in the top left panel). \label{fig:rSFMS}}          
\end{figure*}

\begin{figure*}
\gridline{\fig{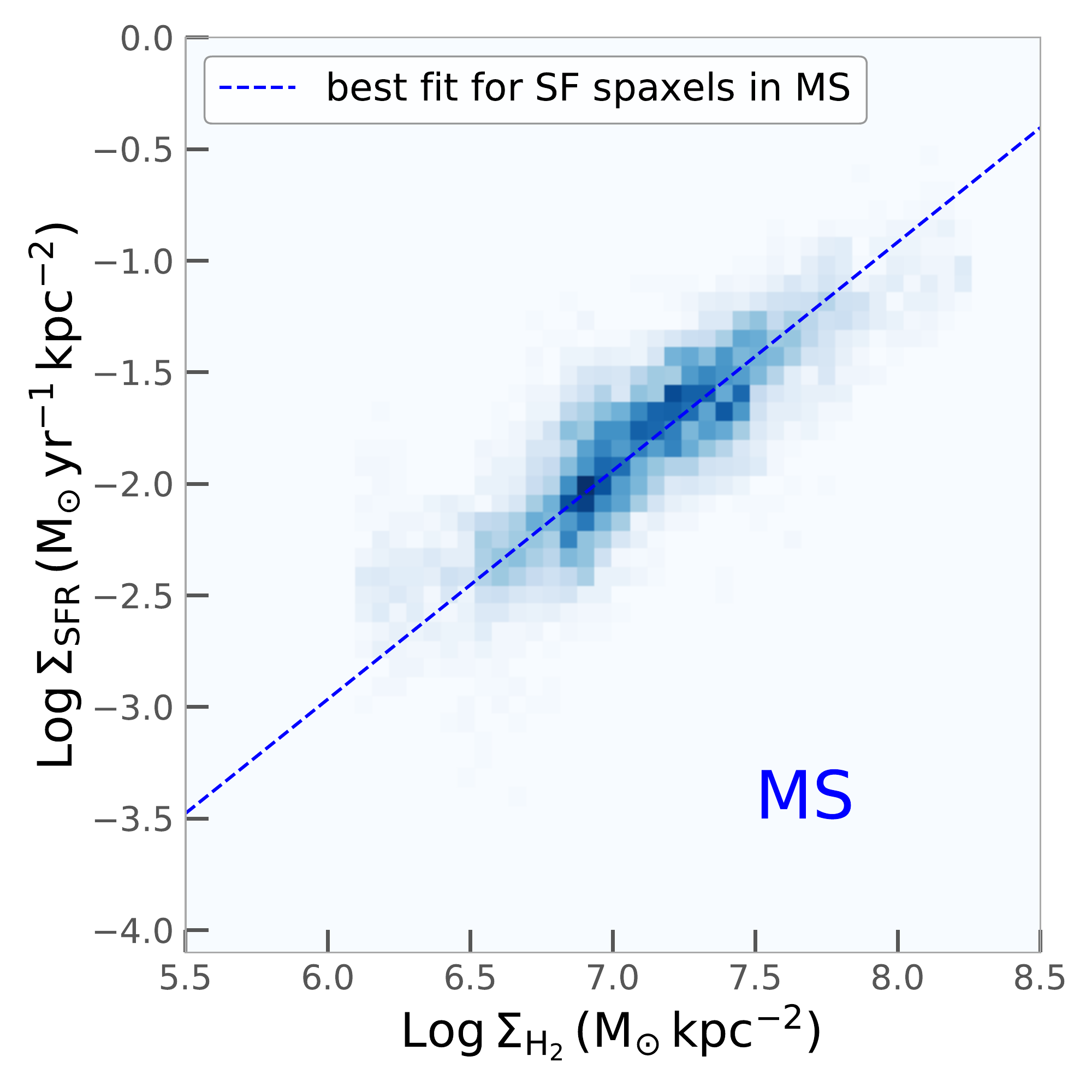}{0.45\textwidth}{(a) SF spaxels in MS galaxies}
          \fig{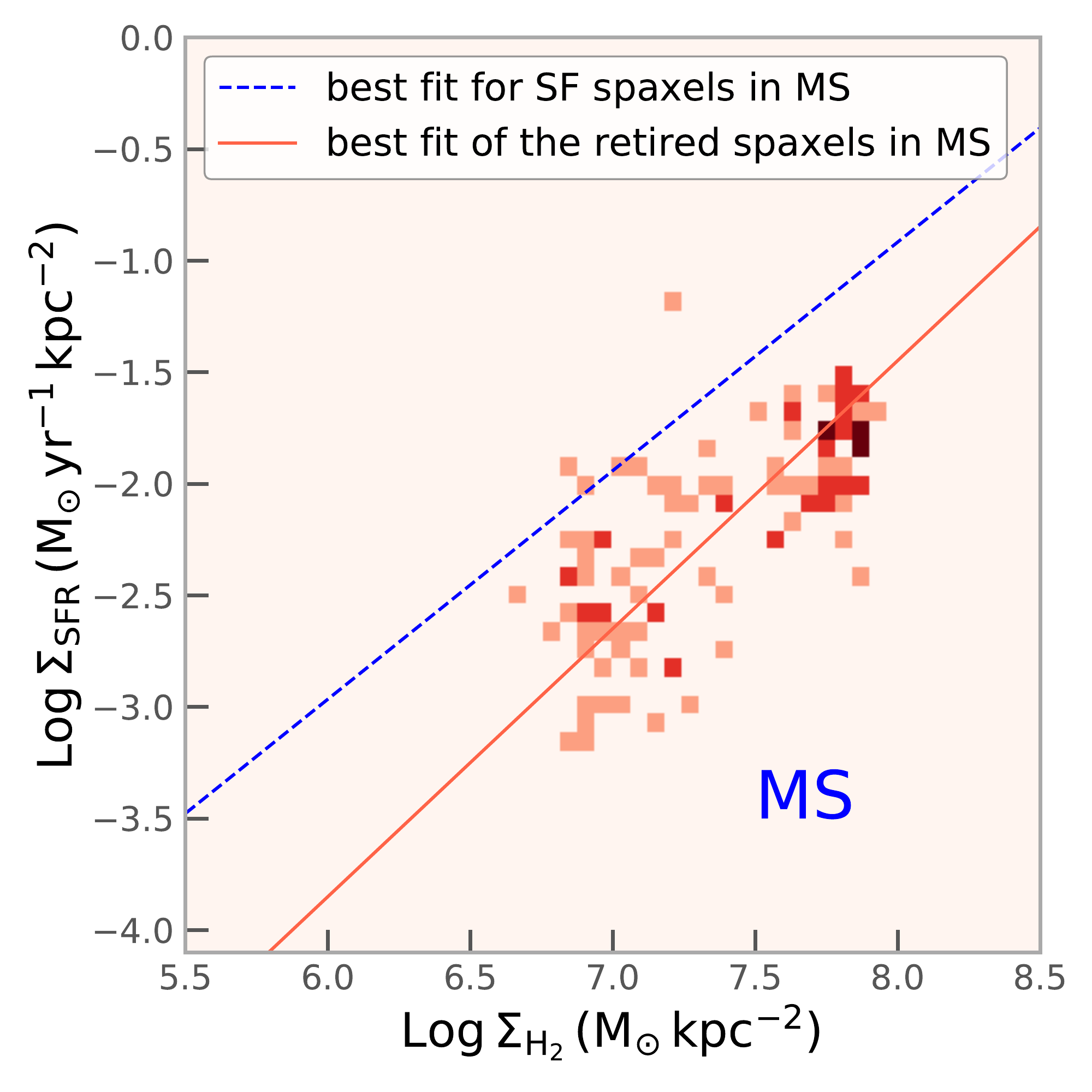}{0.45\textwidth}{(b) Retired spaxels in MS galaxies}}
\gridline{\fig{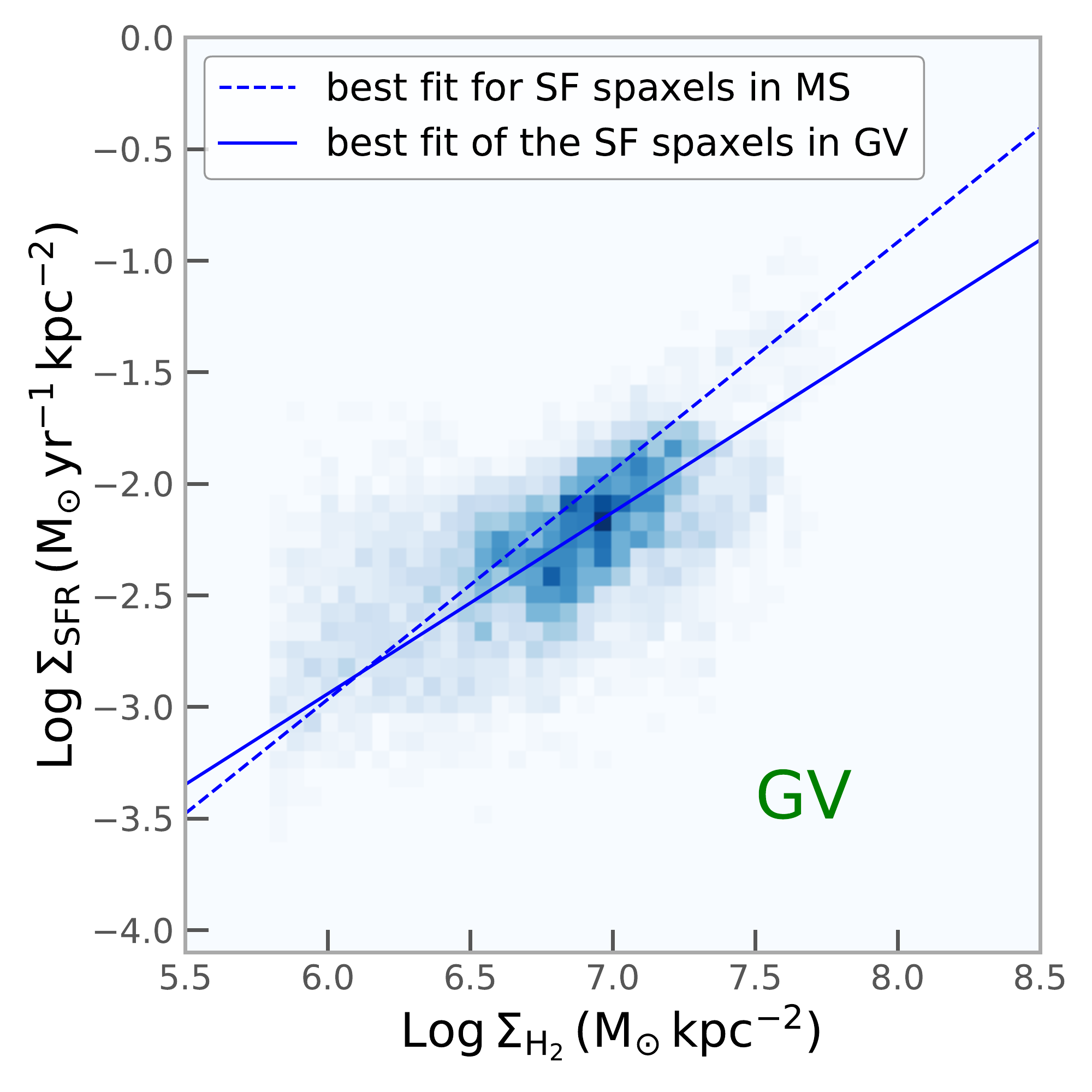}{0.45\textwidth}{(c)  SF spaxels in GV galaxies}
		  \fig{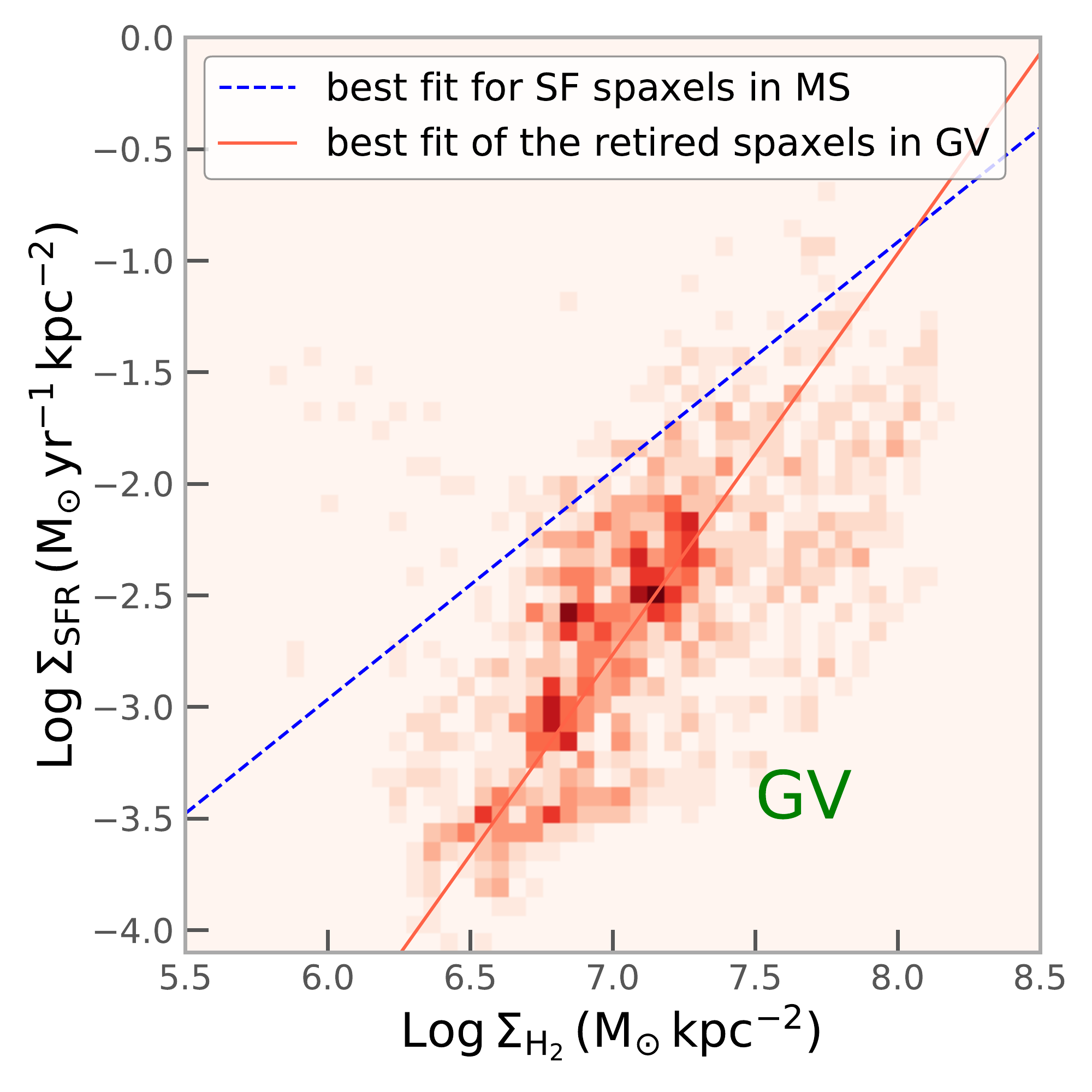}{0.45\textwidth}{(d) Retired spaxels in GV galaxies}}

\caption{Number densities of star-forming (left panels; blue scales) and retired (right panels; red scales) spaxels on the rSK plane for MS (top panels) and GV (bottom panels) galaxies. The solid lines represent the best fit of the data points (blue for star-forming spaxels and red for retired spaxels) shown in each panel. The best-fit parameters are listed in Table \ref{tab:fit}. The blue dashed lines are the same reference lines in each panel, corresponding to the best fit of the rSK derived for the star-froming spaxels in MS galaxies (i.e. a fit to the spaxels in the top left panel).  \label{fig:rSK}}          
\end{figure*}

\begin{figure*}
\gridline{\fig{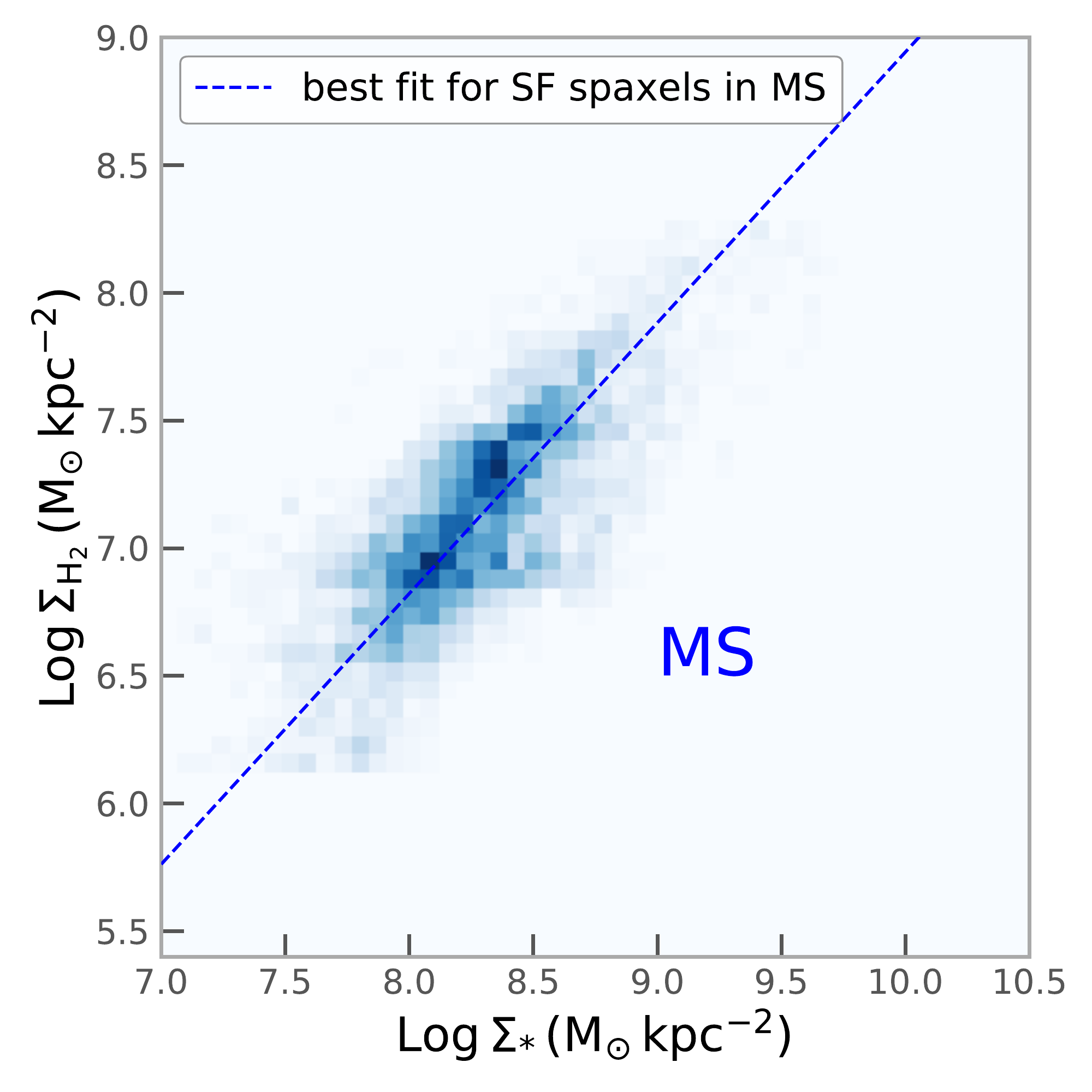}{0.45\textwidth}{(a) SF spaxels in MS galaxies}
          \fig{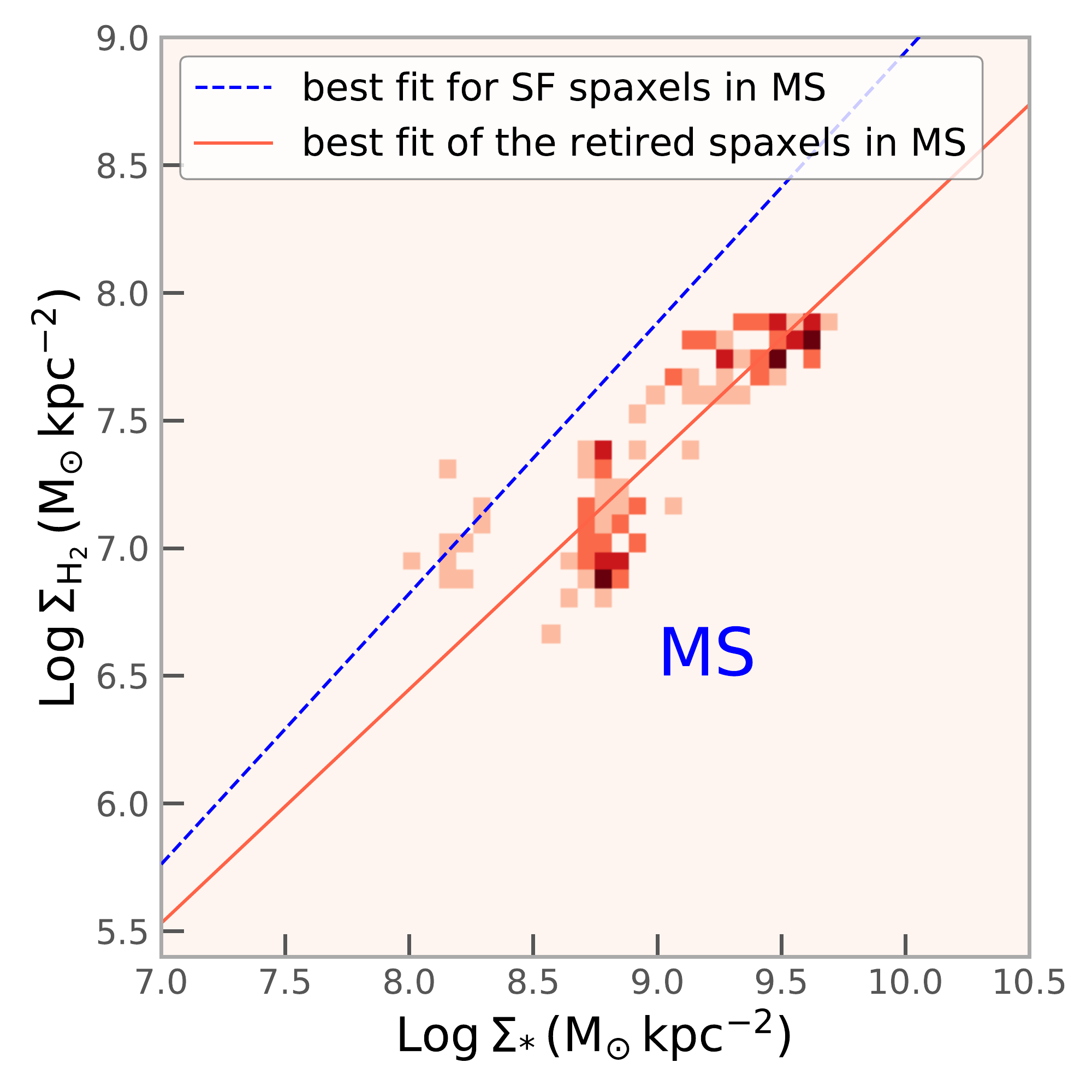}{0.45\textwidth}{(b) Retired spaxels in MS galaxies}}
\gridline{\fig{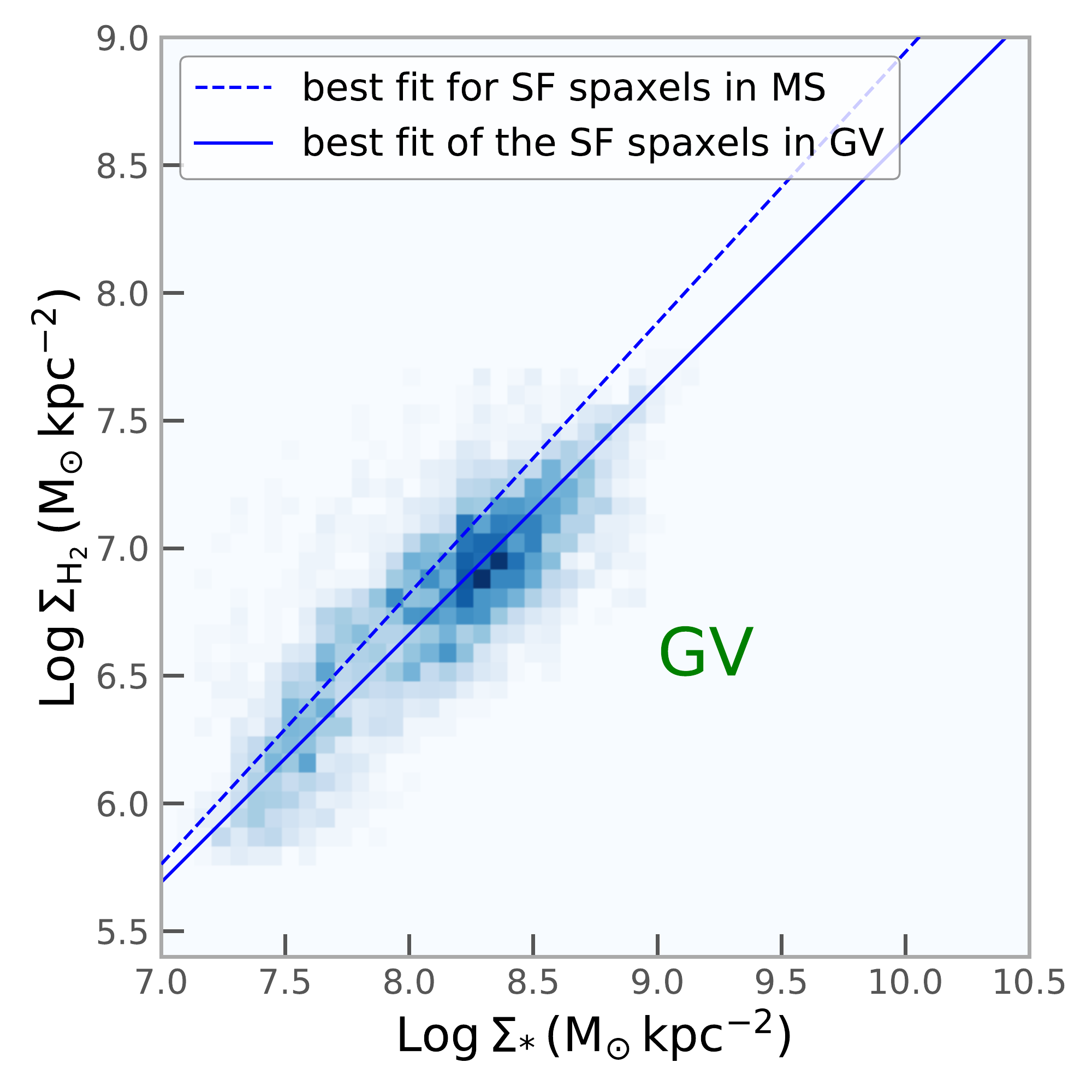}{0.45\textwidth}{(c) SF spaxels in GV galaxies}
		  \fig{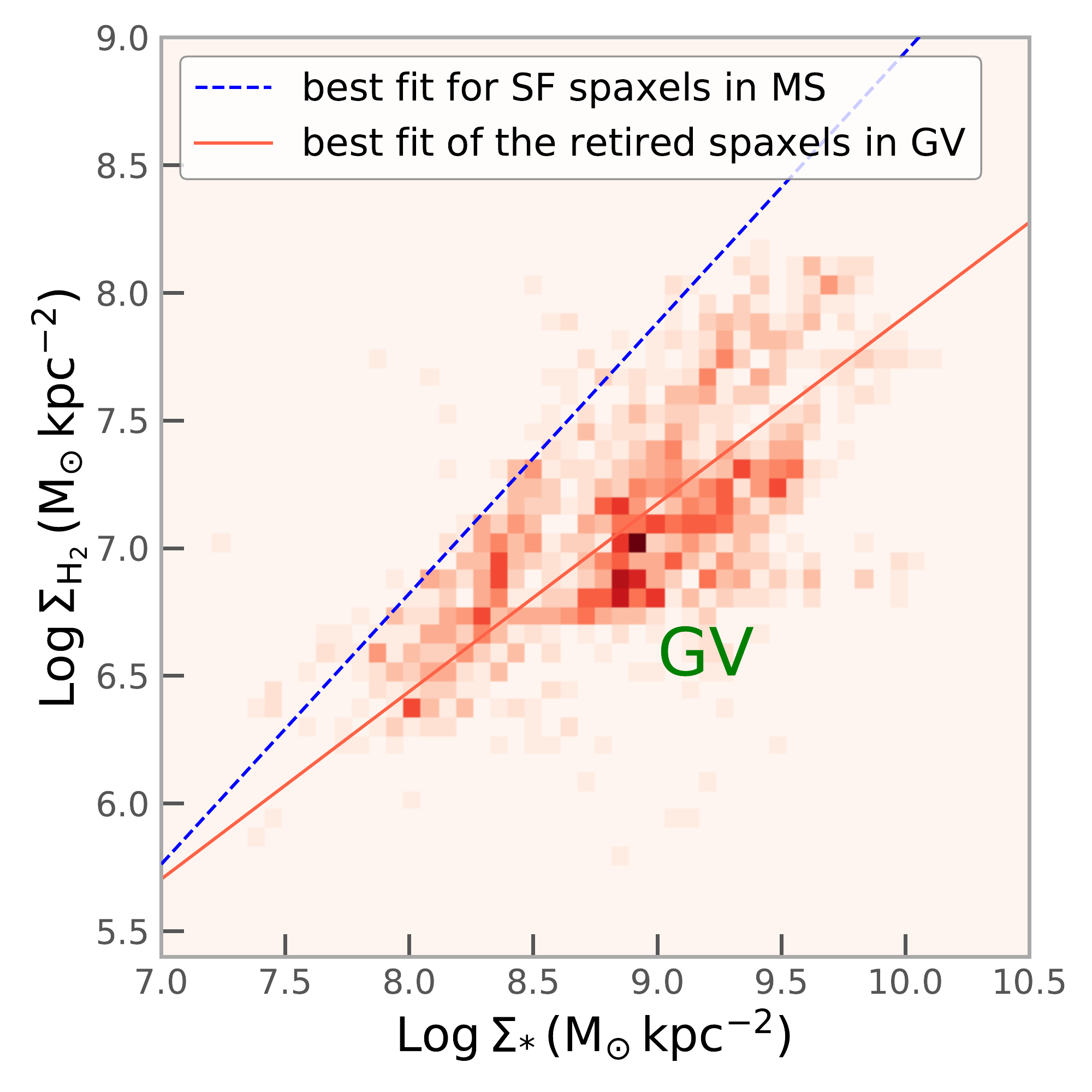}{0.45\textwidth}{(d) Retired spaxels in GV galaxies}}

\caption{Number densities of star-forming (left panels; blue scales) and retired (right panels; red scales) spaxels on the rMGMS plane for MS (top panels) and GV (bottom panels) galaxies. The solid lines represent the best fit of the data points (blue for star-forming spaxels and red for retired spaxels) shown in each panel. The best-fit parameters are listed in Table \ref{tab:fit}. The blue dashed lines are the same reference lines in each panel, corresponding to the best fit of the rMGMS derived for the star-froming spaxels in MS galaxies (i.e. a fit to the spaxels in the top left panel). \label{fig:rMGMS}}          
\end{figure*}

\begin{figure*}
\gridline{\fig{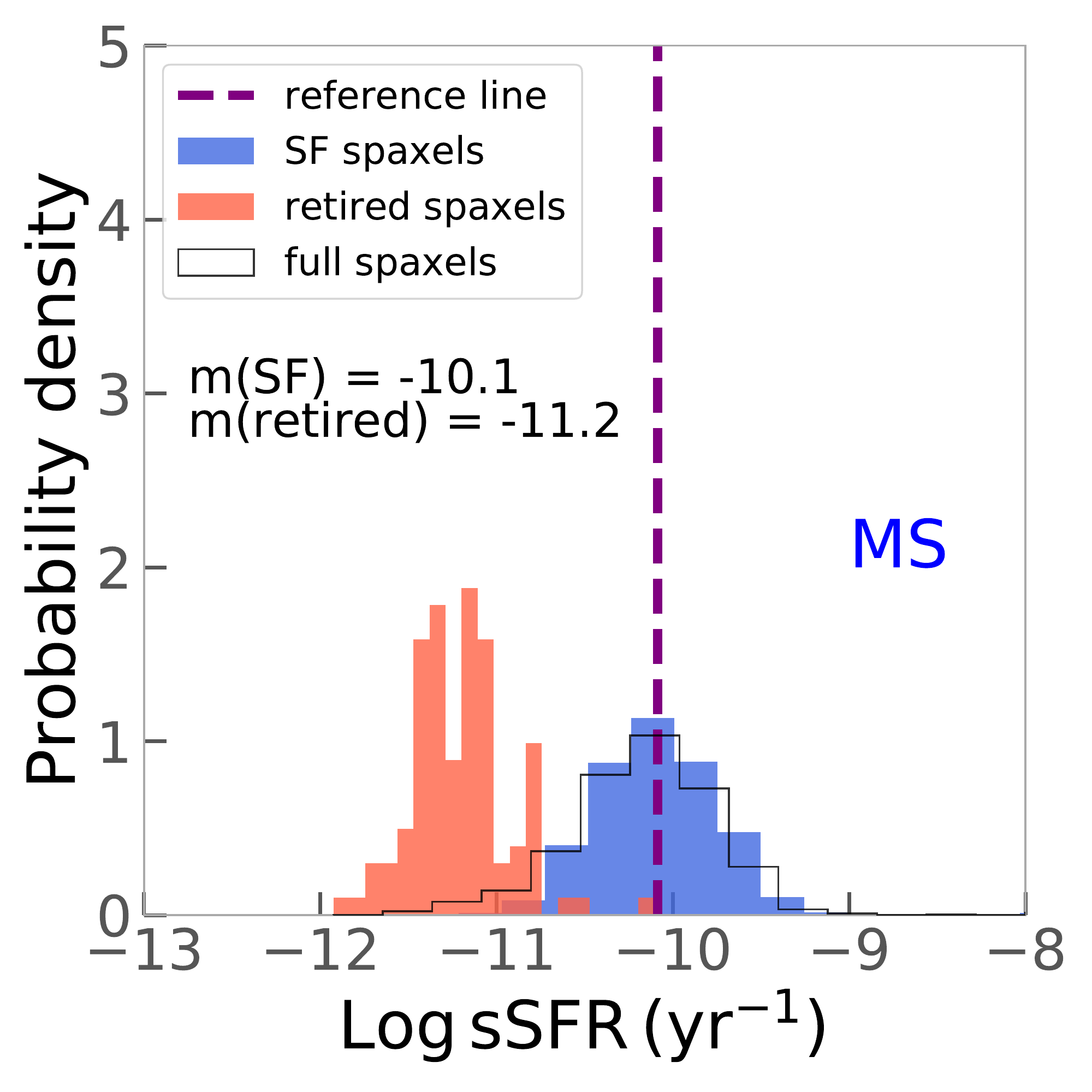}{0.3\textwidth}{}
          \fig{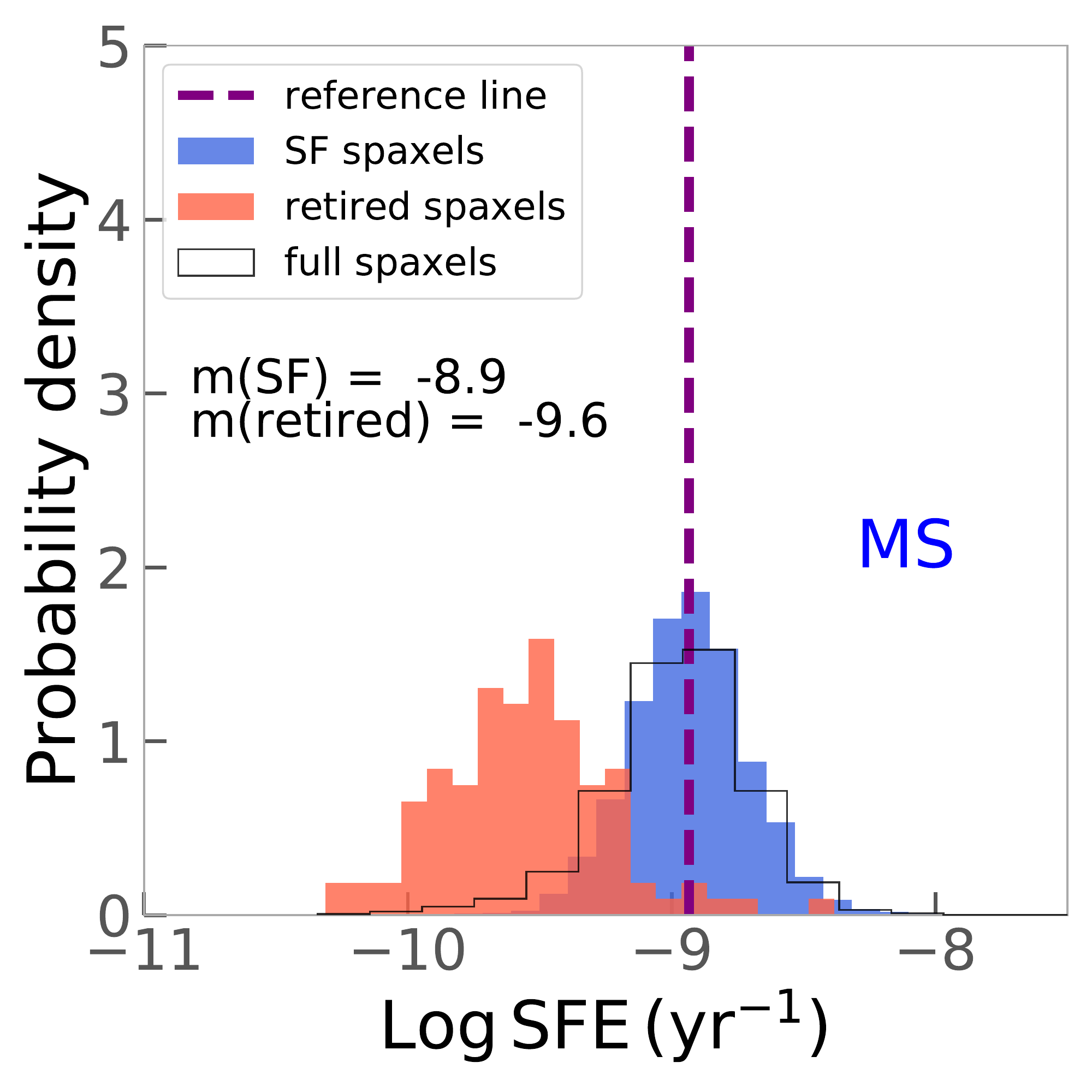}{0.3\textwidth}{}
          \fig{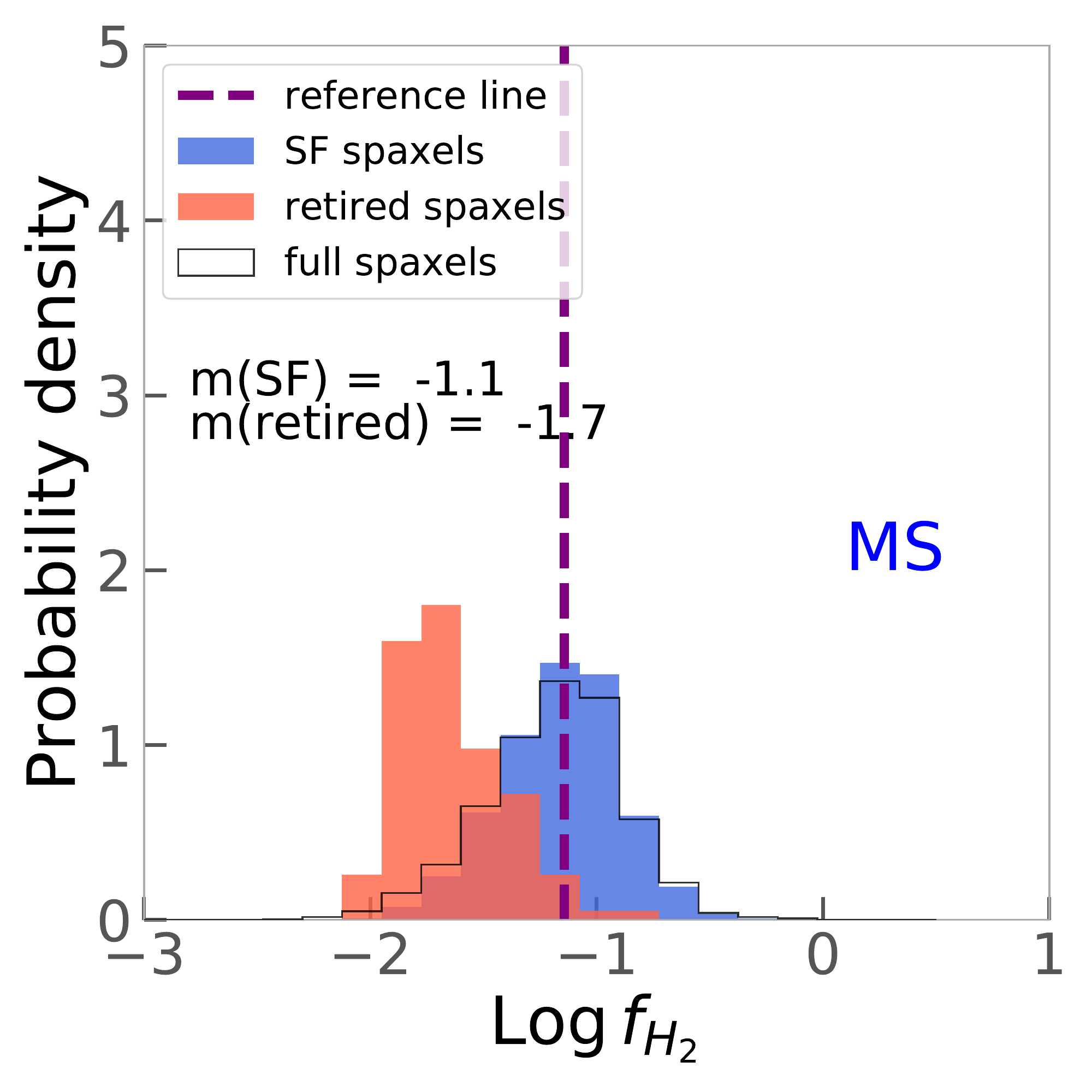}{0.3\textwidth}{}}
\gridline{\fig{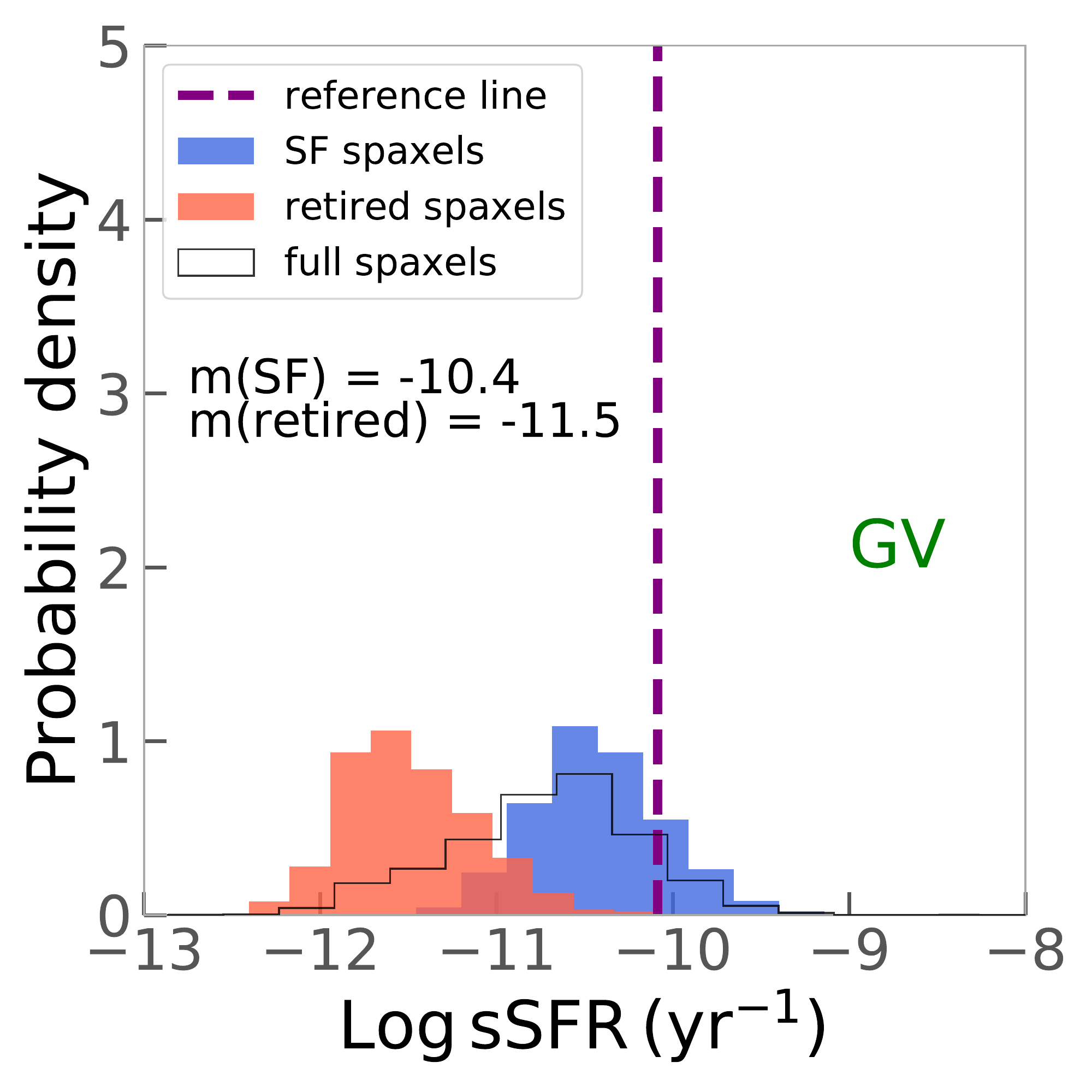}{0.3\textwidth}{}
          \fig{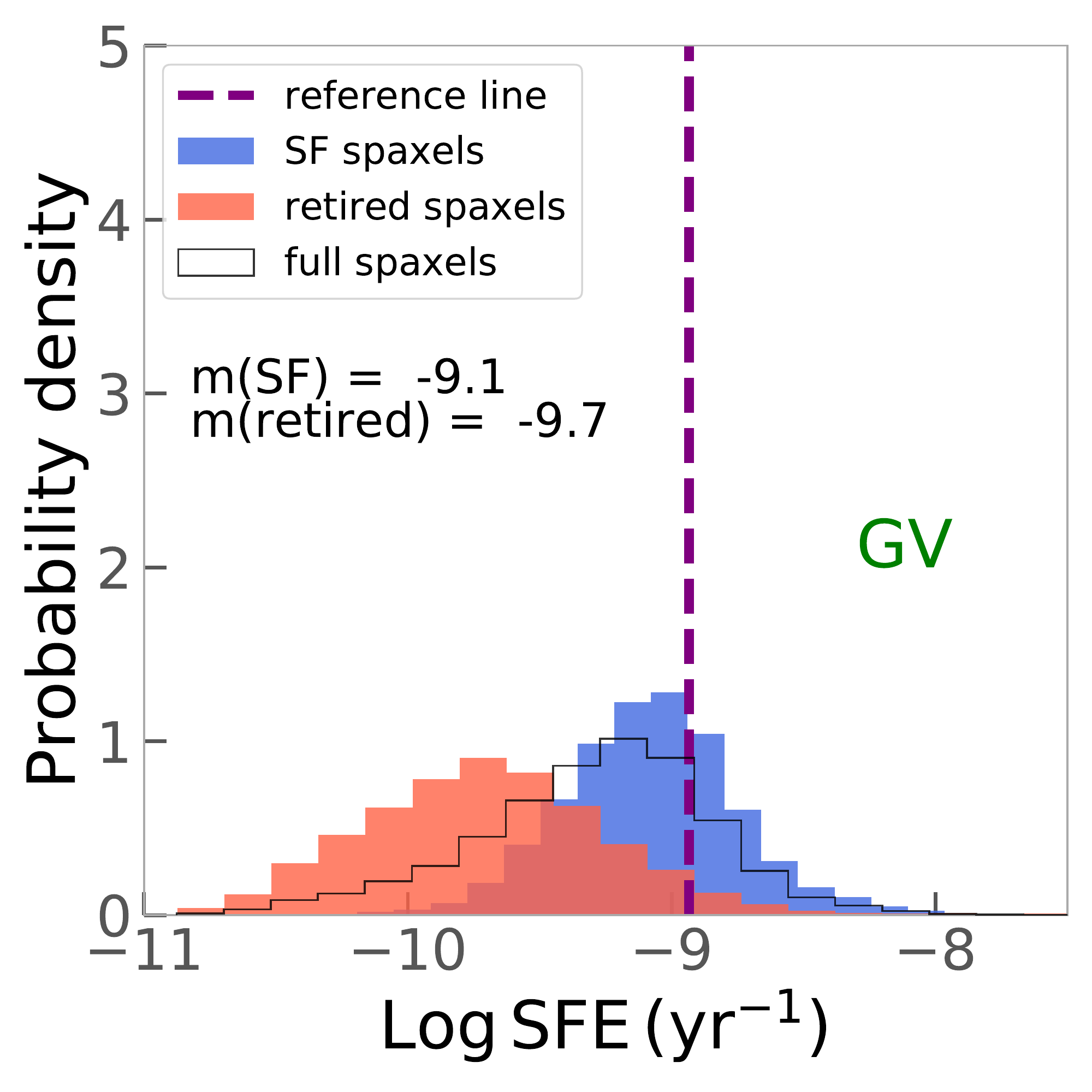}{0.3\textwidth}{}
          \fig{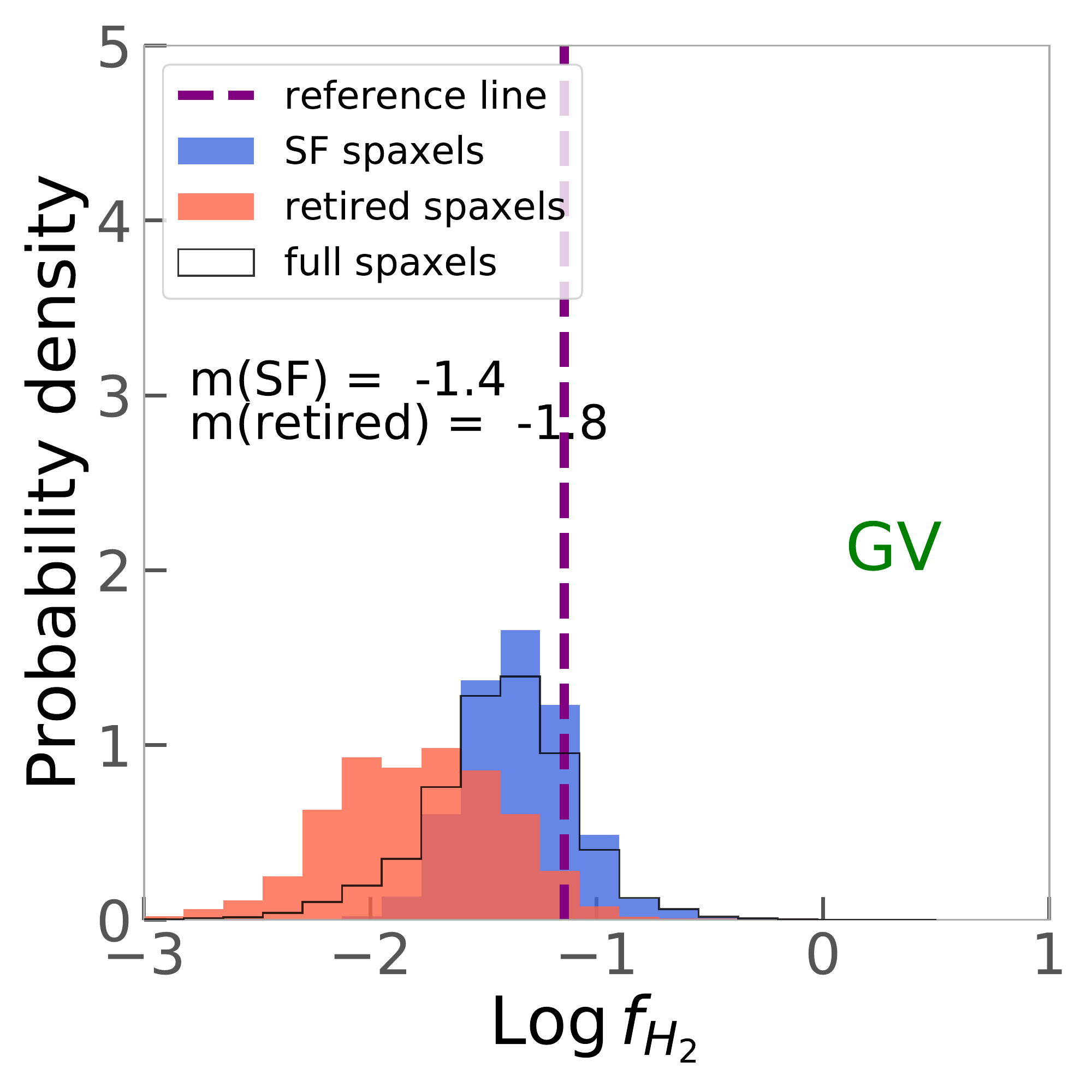}{0.3\textwidth}{}}

\caption{Distributions of sSFR, SFE, and \fh2~for MS (top panels) and GV (bottom panels) galaxies. The probability density distribution for star-forming spaxels is shown in blue histograms, for retired spaxels in red, and for full spaxels in black. Each curve is normalized such that the area under the histogram intergrates to 1. The purple dashed lines correspond to the median values of the quantities of interest measured using the MS star-forming spaxels and are ploted as the same between the MS and GV panels to guide the eyes. The median values of the histograms, $m$ (in log), are reported separately in each panel. \label{fig:his_gas}}          
\end{figure*}

2. The retired spaxels form different sequences from the star-forming spaxels in both GV and MS galaxies on each of the three relations: rSFMS (Figure \ref{fig:rSFMS}), rSK (Figure \ref{fig:rSK}), and rMGMS (Figure \ref{fig:rMGMS}). More specifically, for a given \sigsm, retired spaxels show lower \sigsfr~and lower \sigh2~than star-forming spaxels. For a given \sigh2, retired spaxels show lower \sigsfr~compared to star-forming spaxels. In turn, the sSFR, SFE, and \fh2~of retired spaxels are found to be lower than those of star-forming spaxels by $\sim$ 1.1, 0.6, and 0.5 dex, respectively (Figure \ref{fig:his_gas}). 

3. Compared to the star-forming spaxels in MS galaxies, the star-forming spaxels in GV galaxies show lower values of \sigsfr~and \sigh2~ for a given \sigsm~and show lower \sigsfr~for a given \sigh2 (Figures \ref{fig:rSFMS} -- \ref{fig:rMGMS}). More quantitatively, the sSFR, SFE, and \fh2~of star-forming spaxels in GV galaxies show a reduction by 0.36, 0.14, and 0.21 dex, respectively, when compared with those of star-forming spaxels in MS galaxies (Figure \ref{fig:his_gas}). The Kolmogorov-Smirnov test indicates that the sSFR (as well as SFE or \fh2) between the star-forming spaxels in MS and GV populations are not drawn from the same distribution.

4. When separating a galaxy into bulge and disk regions, we find that SFE is depleted by 0.42$\pm0.10$ (0.35$\pm$0.08) dex  whereas \fh2~is lowered by  0.34$\pm$0.16 (0.25$\pm$0.06) in the bulge (disk) regions of GV galaxies with respect to those of the MS galaxies. This result suggests that SFE and \fh2~can be suppressed in both the inner and outer regions of GV galaxies with respect to the MS galaxies. The significance of the result on the \fh2~suppression in the bulge GV relative to the MS galaxies, however, is only marginal and a larger sample would be required to robustly quantify the reduction of \fh2.

The lower SFE and \fh2~of retired spaxels compared to star-forming spaxels suggest that the cessation of star formation in retired regions is caused by processes that not only reduce \fh2~ but also affect SFE. On the other hand, it is quite intriguing that GV galaxies exhibit lower sSFR, SFE and \fh2~ than MS galaxies even for star-forming spaxels. This suggests that the overall suppressed star formation in GV galaxies with respect to that in MS galaxies is not only associated with the increased fraction of non-star-forming spaxels as illustrated in Figure \ref{fig:his_ion}, but is also related to a globally suppressed sSFR in star-forming regions. This is in line with some previous IFS studies that show a suppressed SFR at all radii for galaxies below the MS \citep{bel18,ell18,san18,med18,wan19}. In a forthcoming paper (H.-A. Pan et al. in prep.), we will examine the radial distribution of the star formation and gas properties in detail for the entire ALMaQUEST sample.

\begin{figure}
\centering
\includegraphics[angle=0,width=0.5\textwidth]{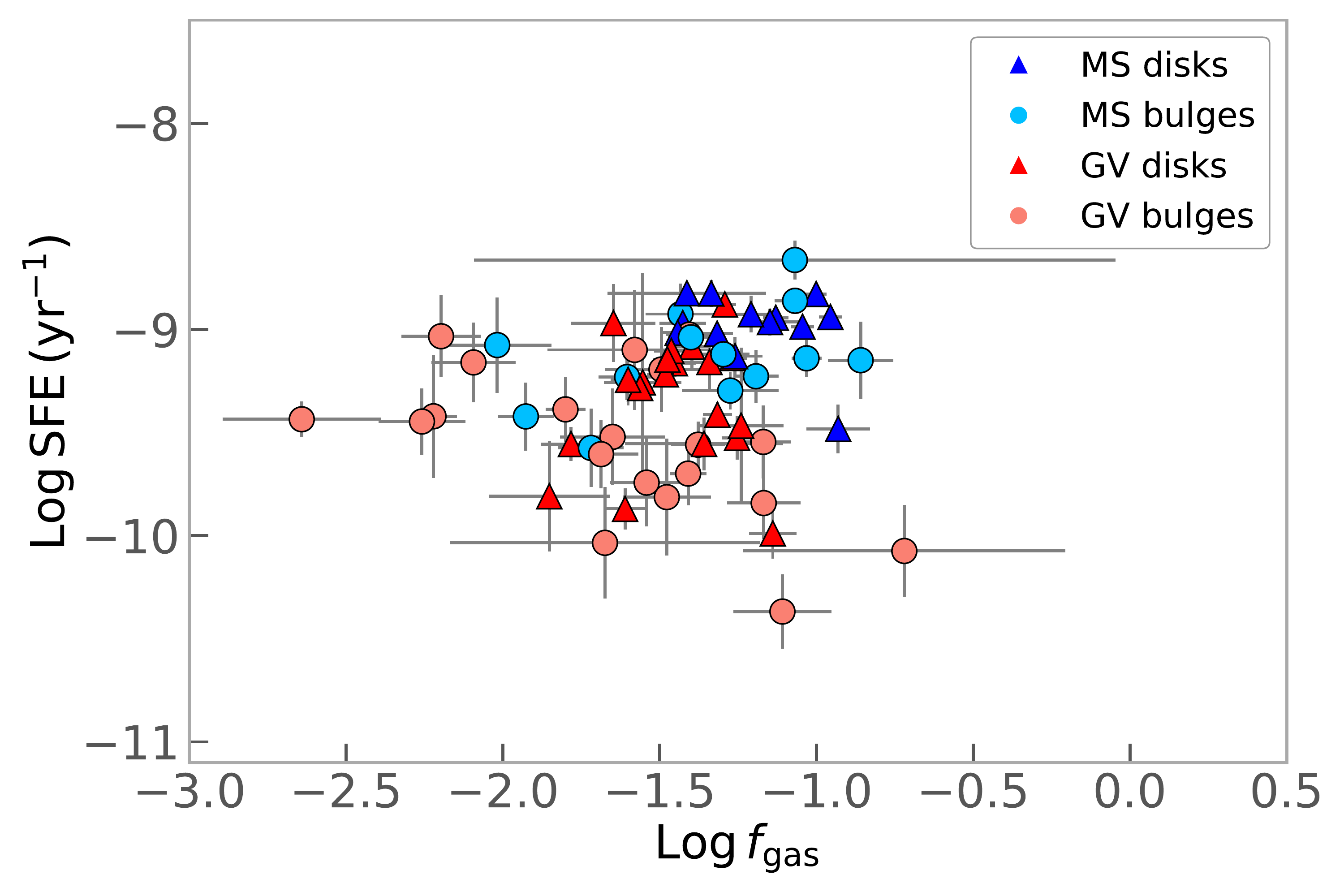}

\caption{ SFE versus \fh2~ for MS (blue colors) and GV (red colors), separately for bulge (circles) and disk (triangles) regions.   \label{fig:bd_sfefgas}}
\end{figure}

In this work, we confirm the picture established in \citet{lin17} that as galaxies move away from the MS toward the quenched state, both the SFE and \fh2~tend to decrease in the disk and bulge regions. Using a sample ten times larger, we are further able to better quantify the effects of SFE and \fh2~in both disk and bulge areas. The relatively lower \fh2~and SFE found in the central regions of galaxies is in agreement with recent studies \citep{lin17,san18, bro20, ell21b}, which favor an inside-out quenching scenario \citep[e.g.,][]{lin19a}. Several processes, including turbulence, magnetic fields, AGN feedback, and the so-called `morphological quenching' may provide pathways to suppress SFE in galaxies \citep{mar09,fed12,fab12}. Not all of them, however, are able to explain the reduced (central) gas fractions at the same time. Among various quenching processes, the most plausible mechanism that is able to account for the depleted gas fraction and SFE is the AGN feedback, which may be able to not only to expel gas away from the central regions, but also to inject the turbulence to suppress star formation \citep{mor17,san18,gar21}. Indeed, \citet{ell21c} have recently demonstrated low central molecular gas fractions in the central regions of AGN host galaxies. Alternatively, it is possible that there may be two types of processes that operate together, one affecting SFE, one removing the gas. Our current analyses on the data, however, are not able to disentangle whether the reduced SFE and \fh2~ share a common source or not, in particular because there is a lack of correlation between SFE and \fh2~ (Figure \ref{fig:bd_sfefgas}).

Many global studies of molecular gas contents have investigated the dependence of the distance from the star-forming main sequence on the gas fraction or SFE \citep[e.g.][]{hua14,sco16,bol17,sai17,tac18,col20,lin20,pio20}. While the relative importance between the gas supply and star formation efficiency is still controversial depending on the choice of the gas tracer and the range of the star formation rate sampled (e.g., within or outside the SFMS), it becomes clear that both the gas fraction and SFE contribute to the change in sSFR to a certain level. Recent resolved studies have also attempted to address this issue by associating the local  sSFR with SFE and \fh2 at (sub)kpc scales \citep{lin17,ell20b,bro20,mor20}. Similarly, it is found that both SFE and \fh2~play roles in regulating the star formation rate above and below the rSFMS, but the relative contributions depend on the properties of host galaxies as well as nature of cold gas. For example, \citet{ell20b} studied the gas content for well-selected star-forming spaxels from the ALMaQUEST galaxies and concluded that SFE is the primary driver determining the scatter of the rSFMS with \fh2~ playing a secondary role. When analyzing several green valley galaxies, \citet{lin17} and \citet{bro20} found that the suppressed star formation in GV galaxies can be attributed to both the deficit in SFE and \fh2~ in either inner or outer regions.  On the other hand, when including the atomic gas, \citet{mor20} found that the total gas fraction has a stronger effect than SFE on the distance from the rSFMS. In this work, we triple the sample size of GV galaxies used in previous studies \citep{lin17,bro20} and find that statistically, the lower SFE in GV galaxies plays a comparable role to the lower \fh2~in reducing the sSFR in GV galaxies. Our results suggest that the suppressed star formation in GV galaxies does not always require a strong depletion in molecular gas. This is in line with some previous global studies that found that SFE is correlated with sSFR or the distance from the MS in the low sSFR regime (i.e., transitioning and retired galaxies) \citep[e.g.,][]{col20,lin20}.

In fact, there are a handful of cases in our sample where the gas fraction is comparable to that of MS but with very low SFE, resulting in low sSFR. One piece of important information currently missing in the current work is the content of dense molecular gas, which is a direct fuel of star formation. In a future paper (L. Lin et al. in prep.), we will report the properties of the dense molecular gas in GV galaxies to understand whether the low SFE is caused by a lack of dense gas (such as HCN or HCO+) or not.
 
\begin{deluxetable*}{lccccccc}
\tabletypesize{\scriptsize}
\tablewidth{0pt}
\tablecaption{Best-fit parameteres ($a$ and $b$) for the three resolved scaling relations (see Equations 1--3) presented in Figure \ref{fig:rSFMS} to Figure \ref{fig:rMGMS} using the orthogonal distance regression (ODR) fitting method. \label{tab:fit}}
\tablehead{
\colhead{Subsample} &
\colhead{$a$ (rSFMS)}&
\colhead{$b$ (rSFMS)}&
\colhead{$a$ (rSK)}&
\colhead{$b$ (rSK)}&
\colhead{$a$ (rMGMS)}&
\colhead{$b$ (rMGMS)}&
\colhead{Number of spaxels}
}
\startdata
 SF spaxels in MS & $1.11 \pm 0.06$ & $-2.13 \pm 0.03$ & $1.02 \pm 0.03$ & $-1.94 \pm 0.01$ & $1.06 \pm 0.04$ & $6.82 \pm 0.02$ &   6122 \\
 SF spaxels in GV & $0.75 \pm 0.05$ & $-2.40 \pm 0.02$ & $0.81 \pm 0.05$ & $-2.13 \pm 0.02$ & $0.97 \pm 0.04$ & $6.66 \pm 0.01$ &   5356 \\
 Retired spaxels in MS & $1.07 \pm 0.32$ & $-3.29 \pm 0.35$ & $1.20 \pm 0.43$ & $-2.65 \pm 0.22$ & $0.92 \pm 0.23$ & $6.45 \pm 0.26$ &    108 \\
 Retired spaxels in GV & $1.25 \pm 0.11$ & $-3.71 \pm 0.11$ & $1.80 \pm 0.22$ & $-2.76 \pm 0.08$ & $0.73 \pm 0.09$ & $6.44 \pm 0.09$ &   1306
\enddata
\end{deluxetable*}

\begin{acknowledgements}

We thank the anonymous referee for his/her helpful comments, which greatly improve the quality of this work. We would also like to thank C. A. L\'opez-Cob\'a for providing a list of barred galaxies used in this work.
This work is supported by the Academia Sinica under
the Career Development Award CDA-107-M03 and the Ministry of Science \& Technology of Taiwan
under the grant MOST 108-2628-M-001-001-MY3. RM acknowledges the ERC Advanced Grant 695671 `QUENCH' and support by the Science and Technology Facilities Council (STFC).

The authors would like to thank the staffs of the East-Asia and North-America ALMA ARCs for their support and continuous efforts in helping produce high-quality data products. This paper makes use of the following ALMA data: ADS/JAO.ALMA\#2015.1.01225.S,  ADS/JAO.ALMA\#2017.1.01093.S, ADS/JAO.ALMA\#2018.1.00541.S, and ADS/JAO.ALMA\#2018.1.00558.S.
ALMA is a partnership of ESO (representing its member states), NSF (USA) and NINS (Japan), together with NRC (Canada), MOST and ASIAA (Taiwan), and KASI (Republic of Korea), in cooperation with the Republic of Chile. The Joint ALMA Observatory is operated by ESO, AUI/NRAO and NAOJ.

Funding for the Sloan Digital Sky Survey IV has been
provided by the Alfred P. Sloan Foundation, the U.S.
Department of Energy Office of Science, and the Participating Institutions. SDSS-IV acknowledges support
and resources from the Center for High-Performance
Computing at the University of Utah. The SDSS web
site is www.sdss.org. SDSS-IV is managed by the Astrophysical Research Consortium for the Participating
Institutions of the SDSS Collaboration including the
Brazilian Participation Group, the Carnegie Institution
for Science, Carnegie Mellon University, the Chilean
Participation Group, the French Participation Group,
Harvard-Smithsonian Center for Astrophysics, Instituto
de Astrof\'isica de Canarias, The Johns Hopkins University, Kavli Institute for the Physics and Mathematics of the Universe (IPMU) / University of Tokyo, Lawrence
Berkeley National Laboratory, Leibniz Institut f\"ur Astrophysik Potsdam (AIP), Max-Planck-Institut f\"ur Astronomie (MPIA Heidelberg), Max-Planck-Institut f\"ur
Astrophysik (MPA Garching), Max-Planck-Institut f\"ur
Extraterrestrische Physik (MPE), National Astronomical Observatory of China, New Mexico State University,
New York University, University of Notre Dame, Observat\'ario Nacional / MCTI, The Ohio State University,
Pennsylvania State University, Shanghai Astronomical
Observatory, United Kingdom Participation Group, Universidad Nacional Aut\'onoma de M\'exico, University of
Arizona, University of Colorado Boulder, University of
Oxford, University of Portsmouth, University of Utah,
University of Virginia, University of Washington, University of Wisconsin, Vanderbilt University, and Yale University. 

\end{acknowledgements}

\end{document}